\documentclass[aps,prd,preprint,showpacs,superscriptaddress,nofootinbib]{revtex4}
\usepackage{epsf,graphics,amssymb,amsfonts,amsmath,array}
\def\bfnabla{\mbox{\boldmath $\nabla$}}

\def\bfsigma{\mbox{\boldmath $\sigma$}}

\def\bfepsilon{\mbox{\boldmath $\epsilon$}}
\def\em{{\rm em}}
\def\lQ{\Lambda_{\rm QCD}}

\def\als{\alpha_{\rm s}}

\def\siml{{\ \lower-1.2pt\vbox{\hbox{\rlap{$<$}\lower6pt\vbox{\hbox{$\sim$}}}}\ }} 
\def\simg{{\ \lower-1.2pt\vbox{\hbox{\rlap{$>$}\lower6pt\vbox{\hbox{$\sim$}}}}\ }}

\def\vbfD{{\ \lower-8pt\vbox{\hbox{\rlap{$\!\leftrightarrow$}\lower8pt\vbox{\hbox{$\!\bf D$}}}}\ }} 
\def\dsl{\,\raise.15ex\hbox{/}\mkern-13.5mu D}

\newcommand{\onec}{1\!\!{\rm l}_c}
 
\newcommand{\nn}{\nonumber}
\newcommand{\be}{\begin{equation}} 
\newcommand{\ee}{\end{equation}}
\newcommand{\bea}{\begin{eqnarray}} 
\newcommand{\eea}{\end{eqnarray}}
\newcommand{\beq}{\begin{equation}}
\newcommand{\eeq}{\end{equation}}
\newcommand{\bqa}{\begin{eqnarray}}
\newcommand{\eqa}{\end{eqnarray}}
\newcommand{\bra}[1]{\ensuremath{\langle#1|}}
\newcommand{\ket}[1]{\ensuremath{|#1\rangle}}
\newcommand{\braQM}[1]{\ensuremath{\langle#1|}}
\newcommand{\ketQM}[1]{\ensuremath{|#1\rangle}}

\newcommand{\Appendix}[1]%
    {%
     \section{#1}%
      }

\begin{document}
\preprint{\vbox{\halign{ &# \hfil\cr & IFUM-841-FT \cr &\today\cr }}}
\title{Model-independent Study of Magnetic Dipole Transitions in Quarkonium}
\author{Nora Brambilla, Yu Jia and Antonio Vairo}
\affiliation{Dipartimento di Fisica dell'Universit\`a di Milano and INFN, via Celoria 16, 20133 Milano, Italy}

\begin{abstract}
We study magnetic dipole (M1) transitions between two quarkonia
in the framework of non-relativistic effective field theories of QCD.
Relativistic corrections of relative order $v^2$ are investigated in a systematic fashion.
Non-perturbative corrections due to color-octet effects are considered for the first time 
and shown to vanish at relative order $v^2$. Exact, all order expressions for the
relevant $1/m$ and $1/m^2$ magnetic operators are derived. 
The results allow us to scrutinize several potential model claims. 
In particular, we show that QCD excludes both contributions to the anomalous
magnetic moment of the quarkonium induced by low-energy fluctuations 
and contributions to the magnetic dipole operators of the type induced by a scalar potential. 
Finally, we apply our results to the transitions $J/\psi\to \eta_c\,\gamma$, 
$\Upsilon(1S) \to \eta_b\,\gamma$, $\Upsilon(2S) \to \eta_b(2S)\,\gamma$, $\Upsilon(2S) \to \eta_b\,\gamma$,
$\eta_b(2S)\to\Upsilon(1S)\,\gamma$, $h_b(1P) \to \chi_{b0,1}(1P)\,\gamma$ and $\chi_{b2}(1P) \to h_b(1P) \,\gamma$ 
by assuming these quarkonium states in the weak-coupling regime. 
Our analysis shows that the $J/\psi\to \eta_c\,\gamma$ width is consistent with a weak-coupling treatment 
of the charmonium ground state, while such a treatment for the
hindered transition $\Upsilon(2S) \to \eta_b\,\gamma$ appears difficult 
to accommodate within the CLEO III upper limit.
\end{abstract}


\maketitle

\newpage

\section{Introduction}
The non-relativistic nature appears to be an essential ingredient to
understand the dynamics of heavy quarkonia. It has been established 
soon after the discovery of the $J/\psi$ in 1974 by many subsequent phenomenological studies on 
numerous observables of the $c\bar c$ and $b\bar b$ bound states. 
Hence, heavy quarkonium is characterized by the interplay among the several supposedly well-separated scales
typical of a non-relativistic system: the heavy quark mass $m$, 
the inverse of the typical size of the quarkonium $1/r \sim mv$ and  the
binding energy $E\sim mv^2$, where $v \ll 1$ is the
velocity of the heavy quark inside the quarkonium. 
Nowadays the effective field theory (EFT) approach has become the paradigm to
disentangle problems with a hierarchy of well-separated scales.
Two effective field theories, non-relativistic QCD (NRQCD) \cite{Caswell:1985ui,Bodwin:1994jh}  and 
potential NRQCD (pNRQCD) \cite{Pineda:1997bj,Brambilla:1999xf}, have been
developed in the last decade. Applications of these two EFTs have 
led to a plethora of new results for several observables in quarkonium physics 
(for a review see \cite{Brambilla:2004jw}).

Among the observables that haven't yet been considered in an EFT framework, are radiative 
transition widths. They have been studied so far almost entirely within phenomenological models 
\cite{Feinberg:1975hk,Sucher:1978wq,Eichten:1978tg,Kang:1978yw,Sebastian:1979gq,
Karl:1980wm,Grotch:1982bi,Moxhay:1983vu,McClary:1983xw,
Grotch:1984gf,Fayyazuddin:1993eb,Lahde:2002wj,Ebert:2002pp,Barnes:2005pb} (a sum rule analysis is provided in 
\cite{Khodjamirian:1979fa}). For a recent review we refer to Eichten's contribution in 
\cite{Brambilla:2004wf}. A textbook presentation can be found in \cite{LeYaouanc:1988fx}. 
Mostly, the models are based on a non-relativistic reduction of some 
relativistic interaction assumed on a phenomenological basis. Eventually, a potential model 
coupled to electromagnetism is recovered. 
In this work, we will describe radiative transitions in the language of EFTs.
In particular, we will employ pNRQCD to study radiative transitions in a model 
independent fashion. 

Two dominant single-photon-transition processes, namely electric dipole (E1) and magnetic 
dipole (M1) transitions, are of considerable interest.
Since, for reasons that will become clear in the following, 
M1 transitions are theoretically much cleaner than E1 transitions,
we will restrict ourselves to M1 transitions in this work.\footnote{
However, E1 transitions are the most copiously observed, because 
their rates are enhanced by $1/v^2$ with respect to the M1 case. 
We will report about E1 transitions elsewhere.}

\begin{figure}
\makebox[-8cm]{\phantom b}
\put(0,5){\epsfxsize=8truecm \epsffile{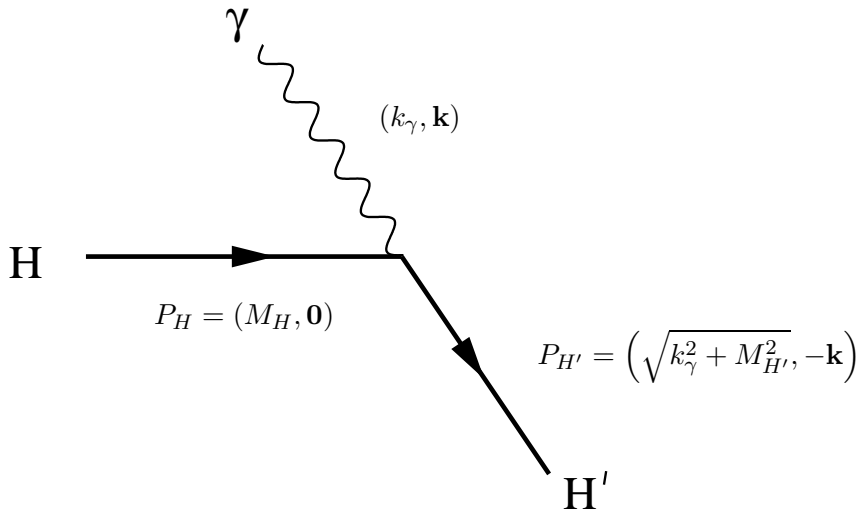}}
\put(55,75){$ P_H = (M_H,{\bf 0})$}
\put(200,60){$ P_{H^\prime} = \left(\sqrt{k_\gamma^2+M_{H^\prime}^2},-{\bf k}\right)$}
\put(140,150){$(k_\gamma,{\bf k})$}
\caption{Kinematics of the radiative transition $H \to H^\prime \gamma$ in the 
rest frame of the initial-state quarkonium $H$. $M_H$ and $M_{H^\prime}$ are the masses of
the initial and final quarkonium, and $k_\gamma = |{\bf k}| = 
(M_H^2-M_{H^\prime}^2)/(2M_H)$ is the energy of the emitted photon.}
\label{figkinematic}
\end{figure}

The kinematics of a transition $H \to H^\prime \gamma$ in the rest frame of $H$,
where $H$ and $H^\prime$ are two quarkonia, is described in Fig.~\ref{figkinematic}. 
In the non-relativistic limit, the M1 transition width between two $S$-wave states is given by
\be
\Gamma_{n^3S_1\to n^\prime{^1S_0}\,\gamma} =  
\frac{4}{3} \, \alpha\, e_Q^2
\, \frac{k^3_\gamma}{m^2}\, \left|\int_0^\infty dr \, r^2 \, R_{n^\prime 0}(r) \, R_{n0}(r) \;
j_0\left( \frac{k_\gamma r}{2}\right)
\right|^2 \,,
\label{M1:NRlimit}
\ee
where  $e e_Q$ is the electrical charge of the heavy quark ($e_b  = -1/3$, $e_c=2/3$), 
$\alpha$ is the fine structure constant and $R_{nl}(r)$ are the radial Schr\"odinger wave functions.
The photon energy $k_\gamma$ is about the difference between the masses of the two
quarkonia, therefore, it is of order 
$mv^2$ or smaller.\footnote{This is in sharp contrast with radiative transitions from a heavy quarkonium to
a light meson, such as $J/\psi\to \eta \gamma$, whereas a hard photon is emitted.} 
Since $r\sim 1/(mv)$, we may expand the spherical Bessel function 
$j_0(k_\gamma r/2) = 1 - (k_\gamma r)^2/24 + \dots$.
At leading order in the multipole expansion, for $n=n'$, 
the overlap integral is 1. Such transitions are usually referred to 
as  {\it allowed}. At leading order, for $n\neq n'$, the overlap integral is 0. 
These transitions  are usually referred to as {\it hindered}. 
The widths of hindered transitions are entirely given by higher-order and
relativistic corrections. 

Equation (\ref{M1:NRlimit}) is not sufficient to explain the observed 
transition widths. In the case of allowed ones, for instance, it overpredicts the observed
$J/\psi\to \eta_c \,\gamma$ transition rate by a factor 2 to 3.
A large anomalous magnetic moment or large relativistic corrections 
have been advocated as a solution to this problem.
Hence, it is crucial to supplement Eq.~(\ref{M1:NRlimit})
with higher-order corrections. EFTs provide a systematic and controlled way for doing it.

EFTs are characterized by a power counting and a range 
of validity (the system must consist of a specific hierarchy of scales).
Errors are controlled by the power counting; higher-order corrections can be systematically included. 
Among these, EFTs include corrections coming from higher-Fock states, typically 
missed in potential models. In particular, in both NRQCD and pNRQCD color-octet 
contributions play a crucial role in some processes.

NRQCD is obtained from QCD by integrating out modes of energy $m$.
The energy scale $m$ is sometimes called {\it hard}.
We denote with $\lQ$ the typical hadronic scale. 
Since $m \gg \lQ$, the matching procedure that ensures the equivalence of the 
two theories may be carried out in perturbation theory. 
At this stage, also hard photons are integrated out. 
However, at the accuracy we are interested in, their contribution is negligible. 

pNRQCD is obtained from NRQCD by integrating out modes of energy $mv$.
This scale is sometimes called {\it soft}. We shall distinguish 
between {\it strongly coupled} quarkonia, for which $mv \sim \lQ$ 
and {\it weakly coupled} quarkonia, for which $mv^2 \simg \lQ$.
In the first case, the matching has to be done in a non-perturbative 
fashion. In the second case, it may be done order by order in the strong-coupling constant.
Low-lying quarkonia are believed to be in the weak-coupling regime, 
higher excitations in the strong-coupling one.
Soft photons are also integrated out at this stage, but its contribution 
is numerically irrelevant with respect to that one coming from soft gluons.
In the strong-coupling regime, the degrees of freedom of pNRQCD (coupled to 
electromagnetism) are singlet quarkonium fields and photons of energy and 
momentum of order $mv^2$ or smaller. The scale $mv^2$ is sometimes called {\it ultrasoft}.
In the weak-coupling regime, there are also octet quarkonium fields and 
ultrasoft gluons. Ultrasoft fields are multipole expanded about the centre-of-mass coordinate.
The power counting of the pNRQCD Lagrangian goes as follows. Ultrasoft gluons and 
virtual photons scale like $mv^2$, the real photon, emitted in a single-photon transition, 
scales like $mv^2$ or smaller. 
In addition, the matching coefficients inherited from NRQCD are series in $\als$.
To simplify the counting, we will assume that $\als(m) \sim v^2$.
In the weak-coupling regime, the matching coefficients of pNRQCD can be calculated 
in perturbation theory. Since the static potential is proportional to
$\als(1/r)/r \sim mv^2$, it follows that $\als(1/r)\sim v$.

In the paper, we will mainly work out pNRQCD in the weak-coupling regime.
Therefore, our final expressions will be applicable only to the lowest quarkonium resonances.
However, some intermediate results will also apply to the strong-coupling
regime. In particular, the $1/m$ and $1/m^2$ matching will be valid to all
orders in $\als$.

Some of the results presented here are new, some may be understood as a rewriting in the language 
of EFTs of results already derived long time ago in the framework of phenomenological models. 
Among others, we will address and answer the following questions.
{\it (i)} What is the size of the quarkonium anomalous magnetic moment? 
{\it (ii)} Is there a scalar interaction contribution to M1 transitions? 
{\it (iii)} What is the size of the octet contributions to M1 transitions?  
We will end up with a rather concise formula which 
takes into account  the full  ${\cal O}( k_\gamma^3\,v^2/m^2)$ relativistic corrections.
We will clarify the validity and range of applicability of 
the widely-used formula of Ref.~\cite{Grotch:1984gf}. Applications to some 
M1 transitions between low-lying quarkonia will be discussed at the end.

The paper is organized as follows. In Sec.~\ref{basic:M1}, we first briefly 
review NRQCD and pNRQCD, then work out the basic 
formalism and calculate the transition widths in the non-relativistic 
limit. In Sec.~\ref{matching:NRQCD:PNRQCD}, we match the electromagnetic interaction 
Lagrangian of pNRQCD relevant for M1 transitions up to $1/m^3$ terms.
In Sec.~\ref{Ov2:RelCorr}, we calculate contributions 
to the transition widths from wave-function corrections and, in particular,
color-octet contributions. In Sec.~\ref{sec:final} we sum all corrections
and give the final formulae valid up to order $k_\gamma^3\,v^2/m^2$. In Sec.~\ref{sec:phen}, the 
decay rates of  $J/\psi \to \eta_c \,\gamma$, $\Upsilon(1S) \to \eta_b \,\gamma$,
$\Upsilon(2S) \to \eta_b(2S) \,\gamma$,  
$\Upsilon(2S) \to \eta_b \,\gamma$, $\eta_b(2S) \to \Upsilon(1S) \,\gamma$, 
$h_b(1P) \to \chi_{b0,1}(1P)\,\gamma$ and $\chi_{b2}(1P) \to h_b(1P) \,\gamma$ 
are calculated. Finally, in Sec~\ref{M1:summary} we conclude.
In one appendix, we discuss alternative ways to derive final-state recoil 
effects, in the other one, issues about gauge invariance.

\section{Magnetic dipole transitions: basic formalism} 
\label{basic:M1}

\subsection{NRQCD}
\label{basic:NRQCD}
NRQCD is the EFT that follows from QCD by integrating out hard modes, i.e. 
modes of energy or momentum of order $m$ \cite{Caswell:1985ui,Bodwin:1994jh}. 
To describe electromagnetic transitions, we need to couple NRQCD to electromagnetism. 
For simplicity, we will call this new EFT also NRQCD. 
The effective Lagrangian is made of operators invariant under the 
$SU(3)_{\rm c}\times U(1)_{\em}$ gauge group.
We display here only the part of the Lagrangian, 
which is relevant to describe M1 transitions at order  $k_\gamma^3\,v^2/m^2$: 
\bea
{\cal L}_{\rm NRQCD} &=& 
\psi^\dagger \left( i D_0 + \frac{{\bf D}^2}{2m} \right) \psi
+ \frac{c_F}{2 m} \psi^\dagger {\bfsigma} \cdot g {\bf B} \psi
- \frac{c_s}{8 m^2} \psi^\dagger \bfsigma \cdot [-i {\bf D}\times, g{\bf E}] \psi  
\nn\\
&& 
+ \frac{c_F^\em}{2 m} \psi^\dagger \bfsigma \cdot e e_Q {\bf B}^{\em} \psi
- \frac{c_S^\em}{8 m^2} \psi^\dagger \bfsigma \cdot [-i{\bf D}\times, e e_Q {\bf E}^{\em}]\psi  
\nn \\ 
&& + \frac{c^\em_{W1}}{8 m^3} \psi^\dagger \{ {\bf D}^2, \bfsigma \cdot e e_Q
    {\bf B}^{\em} \}\psi
- \frac{c^\em_{W2}}{4 m^3} \psi^\dagger 
 {\bf D}^i \bfsigma \cdot e e_Q {\bf B}^{\em} {\bf D}^i \psi
 \nn \\
&& +\frac{c^\em_{p^\prime p}}{ 8 m^3} 
\psi^\dagger \left( \bfsigma\cdot{\bf D} \; e e_Q {\bf B}^\em \cdot {\bf D}
+ {\bf D} \cdot e e_Q {\bf B}^\em \;  \bfsigma\cdot{\bf D} \right) \psi
\nn\\
&& + [\psi \rightarrow i \sigma ^2 \chi^*, A_\mu \rightarrow - A_\mu^T,
      A^\em_\mu \rightarrow - A^{\em}_\mu]
\nn\\
&&
+ {\cal L}_{\rm light} \,,
\label{NRQCD:Lag}
\eea
where 
\be
{\cal L}_{\rm light} =
-\frac{1}{4}F^{\mu\nu\,a}F_{\mu\nu}^a - \frac{1}{4}F^{\mu\nu\,\em}F_{\mu\nu\,\em}
+ \sum_f \bar{q}_f \, i \dsl \,  q_f \,,
\ee
and $\psi$ is the Pauli spinor field that annihilates a heavy quark of mass
$m$, flavor $Q$ and electrical charge $e e_Q$, $\chi$ is the corresponding
one that creates a heavy antiquark, and $q_f$ are the light quark Dirac fields.
The gauge fields with superscript ``$\em$'' are the electromagnetic fields, the 
others are gluon fields, $i D_0 = i \partial_0 - g T^a A^a_0 - e e_Q
A^\em_0$, $i {\bf D} = i \bfnabla + g T^a {\bf A}^a + e e_Q {\bf A}^\em$, 
$[{\bf D} \times, {\bf E}]={\bf D} \times {\bf E} - {\bf E} \times {\bf D}$,  
${\bf E}^i = F^{i0}$, ${\bf B}^i = -\epsilon_{ijk}F^{jk}/2$,  
${\bf E}^{i\,\em} = F^{i0\,\em}$ and  ${\bf B}^{i\,\em} = -\epsilon_{ijk}F^{jk\,\em}/2$ 
($\epsilon_{123} = 1$).

The coefficients $c_F$, $c_S$, $c_F^\em$, $c_S^\em$, $c^\em_{W1}$, $c^\em_{W2}$ and $c^\em_{p^\prime p}$
are the matching coefficients of the EFT. They satisfy some exact relations
dictated by reparameterization (or Poincar\'e) invariance \cite{Manohar:1997qy}:
\bea
c_S^\em &=& 2 c_F^\em -1 \,, \qquad  c_S = 2 c_F -1 \,,  
\label{rpi:cs}
\\
c_{W2}^\em &= & c_{W1}^\em -1\,,
\label{rpi:cw} 
\\
c_{p^\prime p}^\em &= & c_F^\em -1.
\label{rpi:cpp} 
\eea
Note that the $c_{Wi}^\em$ are independent of $c_F^\em$.  
All the coefficients are known at least at one loop \cite{Manohar:1997qy}. 
In particular, we have 
\bea
c_F^\em \equiv 1+ \kappa_Q^\em &=& 1+ C_F \frac{\als}{2\pi} + {\cal O}(\als^2)\,,
\label{kQ} \\
c_{W1}^\em &=& 1+ C_F \frac{\als}{\pi} \left(\frac{1}{12} + \frac{4}{3} \ln \frac{m}{\mu} \right)+ {\cal O}(\als^2)\,,
\eea
where $C_F = (N_c^2-1)/(2N_c) =4/3$ and $C_A=N_c=3$; 
$\kappa_Q^\em$ is usually identified with the anomalous magnetic moment of the heavy quark. 
Since $c_{W1}^\em$ and $c_F^\em$ are $1 + {\cal O}(\als)$, 
$c_{W2}^\em$ and $c_{p^\prime p}^\em$ are ${\cal O}(\als)$. 
$\kappa_Q^\em$ is less than 10\%  for charm and bottom.
One may expect that the magnetic moment of the quarkonium 
may be larger than that, because, apart from inheriting the magnetic moments  
of the quarks, it may get potentially large low-energy contributions.
We will clarify this point in Sec.~\ref{sec:VSB}.

In general, the matching coefficients will contain contributions  
coming from virtual photons of energy or momentum of order $m$, which have
also been integrated out. These contributions are suppressed by powers of $\alpha$ 
and shall be neglected in the following. We will only consider QCD corrections.

\subsection{pNRQCD}
\label{basic:pNRQCD}
NRQCD still contains redundant degrees of freedom in describing a 
quarkonium state far below the open flavor threshold. pNRQCD is the EFT that
follows from NRQCD by further integrating out quarks and gluons  
of momentum and energy of order $mv$  and gluons of
momentum of order $mv$ and energy of order $mv^2$
\cite{Pineda:1997bj,Brambilla:1999xf}.
We consider here pNRQCD in the weak-coupling regime ($mv^2\simg \lQ$).
The degrees of freedom of pNRQCD are quarks of momentum $mv$ and
energy $mv^2$ and (ultrasoft) gluons of energy and momentum of order $mv^2$. 
Since we are interested in quarkonium states, it is convenient to express the
pNRQCD Lagrangian in terms of quark-antiquark fields. These are 
$3 \otimes 3$ tensors in color space and $2 \otimes 2$ tensors in spin space, which 
depend on the centre-of-mass coordinate ${\bf R}$ and the relative distance
${\bf r}$ of the two quarks. At leading order, the pNRQCD Lagrangian very much resembles a
potential model, where the potentials are the matching coefficients of the 
EFT that encode the soft-scale contributions. The pNRQCD Lagrangian, however,
also contains dynamical (ultrasoft) gluons and their interactions with the quark-antiquark fields.

Quarkonium radiative transitions involve real photons of energy and
momentum $k_\gamma$ of order $mv^2$ for hindered transitions and smaller for allowed ones.
These transitions are described by pNRQCD if photons of momentum $mv$ are integrated out from NRQCD and 
photons of energy and momentum of order $mv^2$ or lower are explicitly coupled 
to the quark fields in the pNRQCD Lagrangian.
To ensure that gluons and photons are of energy and momentum not larger than 
$mv^2$, all gauge fields are multipole expanded in the relative distance ${\bf
  r}$, and, therefore, depend on the centre-of-mass coordinate ${\bf R}$ only.
 
Gauge invariance can be made manifest at the Lagrangian level by 
reexpressing the quark-antiquark fields in terms of fields 
that transform like singlets under $U(1)_{\em}$ and 
like singlets or octets under $SU(3)_{\rm c}$ gauge transformations.
We denote these fields as ${\rm S} = S
\onec/\sqrt{N_c}$ and ${\rm O} = \sqrt{2}\, O^aT^a$, respectively.

The pNRQCD Lagrangian, which is relevant to describe M1 transitions at order
$k_\gamma^3\,v^2/m^2$, is given by 
\bea
{\cal L}_{\rm pNRQCD} &=& \int d^3r \;  {\rm Tr} \, \Bigg\{ 
{\rm S}^\dagger 
\left( i\partial_0 + \frac{\bfnabla^2}{4 m} +\frac{\bfnabla_r^2}{m}  - V_S
\right) {\rm S} 
\nn \\
&&  \hspace{16mm}
+ {\rm O}^\dagger \left( iD_0 + \frac{{\bf D}^2}{4 m} +
  \frac{\bfnabla_r^2}{m} - V_O
\right) {\rm O}
\nn \\
&& \hspace{16mm}
+ V_A\;({\rm O}^\dagger {\bf r}\cdot g{\bf E} \,{\rm S} + {\rm S}^\dagger {\bf r}\cdot g{\bf E} \,{\rm O})
+ V_B \, \frac{ \left\{ {\rm O}^\dagger , {\bf r}\cdot g{\bf E} \right\}}{2}{\rm O} \Bigg\}
\nn\\
&& + {\cal L}_{\gamma\, \rm pNRQCD}
\nn\\
&& + {\cal L}_{\rm light}\,,
\label{pNRQCD:Lag}
\eea 
where 
\bea
{\cal L}_{\gamma\, \rm pNRQCD} &=& \int d^3 r \;  {\rm Tr} \, \Bigg\{ 
V_A^\em \; {\rm S}^\dagger {\bf r}\cdot e e_Q {\bf E}^{\em} {\rm S} 
\nn \\
&& \hspace{16mm}
+ \frac{
1}{2 m}
\; V^{\frac{\sigma\cdot B}{m}}_S 
\; \left\{{\rm S}^\dagger , \bfsigma \cdot e e_Q {\bf B}^{\em}\right\} {\rm S} 
\nn \\
&& \hspace{16mm}
+ \frac{
1}{16 m}
\; V^{(r\cdot \nabla)^2 \frac{\sigma\cdot B}{m}}_S
\; \left\{ {\rm S}^\dagger , {\bf r}^i {\bf r}^j 
    (\bfnabla^i \bfnabla^j \bfsigma \cdot e e_Q {\bf B}^{\em})\right\} {\rm S} 
\nn \\
&& \hspace{16mm}
+ \frac{
1
}{2 m}
\; V^{\frac{\sigma\cdot B}{m}}_O
\; \left\{{\rm O}^\dagger , \bfsigma \cdot e e_Q {\bf B}^{\em} \right\} {\rm O} 
\nn \\
&& \hspace{16mm}
+ \frac{1}{4 m^2}
\; \frac{V^{\frac{\sigma\cdot (r \times r \times B)}{m^2}}_{S}}{r} 
\; \left\{{\rm S}^\dagger , \bfsigma\cdot\left[ \hat{\bf r} \times  \left(
  \hat{\bf r}\times e e_Q {\bf B}^{\em} \right) \right] \right\} {\rm S} 
\nn \\
&& \hspace{16mm}
+ \frac{1}{4 m^2}
\; \frac{V^{\frac{\sigma\cdot B}{m^2}}_{S}}{r} 
\; \left\{{\rm S}^\dagger , \bfsigma \cdot e e_Q {\bf B}^{\em}\right\} {\rm S} 
\nn \\
&& \hspace{16mm}
- \frac{
1}{16 m^2} 
\; V^{\frac{\sigma\cdot \nabla \times E}{m^2}}_S
\; \left[{\rm S}^\dagger,  \bfsigma \cdot \left[-i\bfnabla \times, e e_Q {\bf E}^\em \right]\right] {\rm S}
\nn \\
&& \hspace{16mm}
- \frac{
1}{16 m^2} 
\; V^{\frac{\sigma\cdot \nabla_r \times r\cdot \nabla E}{m^2}}_S
\; \left[ {\rm S}^\dagger,  \bfsigma \cdot \left[-i\bfnabla_r \times, 
{\bf r}^i (\bfnabla^i e e_Q  {\bf E}^\em) \right]\right] {\rm S}
\nn \\
&& \hspace{16mm}
+ \frac{
1}{4 m^3} 
\; V^{\frac{\nabla_r^2 \, \sigma\cdot B}{m^3}}_S
\; \left\{ {\rm S}^\dagger , \bfsigma \cdot e e_Q {\bf
  B}^{\em} \right\} \bfnabla_r^2 {\rm S} 
\nn \\
&& \hspace{16mm}
+ \frac{
1}{4 m^3} 
\; V^{\frac{(\nabla_r\cdot\sigma)\, (\nabla_r\cdot B)}{m^3}}_S
\; \left\{ {\rm S}^\dagger , \bfsigma^i \, e e_Q {\bf
  B}^{\em\,j} \right\} \bfnabla_r^i\bfnabla_r^j {\rm S} \Bigg\}\,.
\label{gammapNRQCD:Lag}
\eea
If not differently specified, all gauge fields are calculated in the centre-of-mass coordinate ${\bf R}$, 
$i D_0 {\rm O} = i \partial_0 {\rm O} - g [T^a A^a_0, {\rm O}]$, 
$i {\bf D} {\rm O} = i \bfnabla {\rm O} + g [T^a {\bf A}^a, {\rm O}]$, 
$\bfnabla^i = \partial/\partial {\bf R}^i$ and $\bfnabla_r^i = \partial/\partial {\bf r}^i$.
The trace is over color and spin indices.

In the initial quarkonium rest frame, the power counting goes as
follows:\footnote{
For simplicity, we give the power counting in the case $mv^2\sim \lQ$ only.}
$\bfnabla_r \sim mv$, $r\sim 1/mv$ and 
${\bf E}, {\bf B} \sim m^2 v^4$. The electromagnetic fields associated to the external 
photon scale like ${\bf E}^\em, {\bf B}^\em \sim k_\gamma^2$. The centre-of-mass derivative 
$\bfnabla$ acting on the recoiling final quarkonium state or emitted photon 
is of order $k_\gamma$. Operators that have not been 
displayed are suppressed either in the power counting (e.g. $1/m^4$ singlet operators)
or in the matching coefficients (e.g. a $1/(m^3 \,r^2) 
\{{\rm S}^\dagger , \bfsigma \cdot e e_Q {\bf B}^{\em}\} {\rm S}$  operator 
does not show up at tree level) or because they project on higher-order Fock states 
(e.g. $1/m^2$ octet operators).

The coefficients $V$ in Eqs.~(\ref{pNRQCD:Lag}) and (\ref{gammapNRQCD:Lag}) are the matching coefficients 
of pNRQCD. The matching coefficients of Eq.~(\ref{pNRQCD:Lag}) have been calculated 
in the past years. We refer the reader to \cite{Brambilla:2004jw} and references therein.
In the following, we will calculate the matching coefficients of Eq.~(\ref{gammapNRQCD:Lag}). 
Here, we only note that since $mv \gg \lQ$, they may be calculated in perturbation theory. 
$V_S$ and $V_O$ play the role of a singlet and octet potential.
They may be arranged in powers of $1/m$.   
The static contribution is the Coulomb potential:
\be
V_S^{(0)} = - C_F \frac{\alpha_{V_S}}{r}, 
\qquad 
V_O^{(0)} = \frac{1}{2N_c} \frac{\alpha_{V_O}}{r},
\label{static:potential}
\ee
where, at leading order, $\alpha_{V_S} = \alpha_{V_O} = \als$. 
In a Coulombic system $\als(1/r) \sim v$.

Let us discuss the different terms appearing in Eqs. (\ref{pNRQCD:Lag}) and (\ref{gammapNRQCD:Lag}).
The first three lines of Eq.~(\ref{pNRQCD:Lag}) display the pNRQCD Lagrangian in the limit of zero coupling
to the photons. The third line describes the coupling of the quarkonium fields to ultrasoft gluons at order ${\bf
  r}$ in the multipole expansion. Higher-order terms are irrelevant for
the present purposes. At tree level, the coefficients $V_A$ and $V_B$ are equal to 1. 
Equation (\ref{gammapNRQCD:Lag}) provides the part of the pNRQCD interaction Lagrangian 
coupled with the electromagnetic field relevant for M1 transitions.
The first line describes the coupling of the quarkonium singlet field to
ultrasoft photons at order ${\bf r}$. 
This is the familiar E1 transition operator ($V_A^\em =1$ at tree level). 
As we will discuss in the following, if the recoiling of the final-state quarkonium  is taken into account, this
term contributes to M1 transitions. A similar term involving the 
coupling with the octet field is suppressed in the transition amplitude. 
From the second line on, we display spin-dependent operators coupled to ultrasoft photons. Lines two and three
come from multipole expanding the magnetic dipole operator at 
${\cal O}(r^0)$ and ${\cal O}(r^2)$ respectively. The order ${\bf r}$ term 
does not contribute to M1 transitions. Line four represents the leading
magnetic dipole operator for octet quarkonium fields.

\subsection{Radiative transitions}
\label{radiative:transitions}
The process $H \to H^\prime \gamma$ is described in the rest frame of the initial quarkonium state
$H$ by the kinematics of Fig.~\ref{figkinematic}. The transition width is given by:
\bea
\Gamma_{H \to H^\prime \gamma} &=& \int \frac{d^3P^\prime}{(2\pi)^3} \frac{d^3k}{(2\pi)^3}\,
\frac{1}{2k}
(2\pi)^4\delta^4(P_H-k-P^\prime)\, 
\nn\\
&& \hspace{40mm}
\times
\frac{1}{N_\lambda}\sum_{\lambda\lambda^\prime\sigma}
\big| {\cal A}\left[H({\bf 0},\lambda) \to H^\prime({\bf P}^\prime,\lambda^\prime) 
\gamma({\bf k},\sigma)\right] \big|^2
\nn\\
&=& \frac{1}{8\pi^2} \left( 1 - \frac{k_\gamma}{M_H} \right) \int_0^\infty \!\!\! dk \,  k\; 
\int \!\! d\Omega(\hat{\bf k}) \;
\delta(k-k_\gamma)\, 
\nn\\
&& \hspace{40mm}
\times
\frac{1}{N_\lambda}\sum_{\lambda\lambda^\prime\sigma}
\big| {\cal A}\left[H({\bf 0},\lambda) \to H^\prime(-{\bf k},\lambda^\prime) 
\gamma({\bf k},\sigma)\right] \big|^2,
\label{transitionwidth}
\eea
where $P^{\prime\,\mu} = (\sqrt{{\bf P}^{\prime\,2} + M_{H^\prime}^2},{\bf P}^\prime)$, 
$k^\mu = (|{\bf k}|,{\bf k})$ and 
\bea
{\cal A}\left[H({\bf 0},\lambda) \to H^\prime(-{\bf k},\lambda^\prime) \gamma({\bf k},\sigma)\right] 
\; (2\pi)^3\delta^3({\bf P}^\prime + {\bf k})
& & 
\nn\\
& & \hspace{-35mm}
= - \bra{ H^\prime({\bf P}^\prime,\lambda^\prime) \,\gamma({\bf k},\sigma)}
\int d^3R\, {\cal L}_{\gamma\,\rm pNRQCD}\ket{H({\bf 0},\lambda)} 
\label{transitionamplitude}
\,.
\eea
In Eq.~(\ref{transitionwidth}), the initial state is averaged over the polarizations, whose number is $N_\lambda$.

The quarkonium state $\ket{H({\bf P},\lambda)}$ is an eigenstate 
of the pNRQCD Hamiltonian with the quantum numbers of the quarkonium $H$.
It has the non-relativistic normalization: 
\be
\bra{H({\bf P}^\prime,\lambda^\prime)} H({\bf P},\lambda) \rangle =
\delta_{\lambda\lambda^\prime} \,(2\pi)^3\delta^3({\bf P}-{\bf P}^\prime)\,.
\label{statenormalization}
\ee
The photon state $\ket{\gamma({\bf k},\sigma)}$ is normalized in the 
usual Lorentz-invariant way:
\be
\bra{\gamma({\bf k},\sigma)} \gamma({\bf k}',\sigma')\rangle = 2k \, \delta_{\sigma\sigma'}\, 
(2\pi)^3\delta^3({\bf k}-{\bf k}^\prime)\,.
\ee

\subsection{Quarkonium states}
\label{secQS}
According to the power counting, the leading-order pNRQCD Hamiltonian is given
by
\bea
H^{(0)}_{\rm pNRQCD} &=& \int d^3R \int d^3r \, {\rm Tr}\, \left\{
{\rm S}^\dagger \, h_S^{(0)} \, {\rm S} + 
{\rm O}^\dagger \, h_O^{(0)} \, {\rm O} \right\}+ H_{\rm light}\,,
\eea
where 
\be
 h_S^{(0)} = - \frac{\bfnabla^2_r}{m} + V_S^{(0)}\,,
\qquad\qquad
 h_O^{(0)} = - \frac{\bfnabla^2_r}{m} + V_O^{(0)}\,,
\ee
and $H_{\rm light}$ is the Hamiltonian that corresponds to ${\cal L}_{\rm light}$.
The spectrum of pNRQCD has been first studied in \cite{Brambilla:1999xf}, to
which we refer for discussions. We call $\ket{H({\bf P},\lambda)}^{(0)}$ the
subset of eigenstates made by a quark-antiquark pair in a singlet representation:
\be
\ket{H({\bf P},\lambda)}^{(0)} = \int d^3R \int d^3r \; e^{i{\bf P}\cdot {\bf R}} \, {\rm Tr} \left\{
\Phi^{(0)}_{H(\lambda)}({\bf r})\,{\rm S}^\dagger({\bf r},{\bf R})\ket{0}\right\}\,,
\label{ketHP00}
\ee
where $\ket{0}$ is a state that belongs to the Fock subspace containing no heavy quarks 
but  an arbitrary number of ultrasoft gluons, photons and light quarks. 
The state $\ket{0}$ is normalized in such a way that Eq.~(\ref{statenormalization}) is fulfilled.
The function $\Phi^{(0)}_{H(\lambda)}({\bf r}) = 
\bra{0}{\rm S}({\bf r},{\bf R})\ket{H({\bf 0},\lambda)}^{(0)}$ is an
eigenstate of the spin and orbital angular momentum of the quarkonium and 
satisfies the Schr\"odinger equation
\be
h_S^{(0)}\,\Phi^{(0)}_{H(\lambda)}({\bf r}) = E^{(0)}_H \Phi^{(0)}_{H(\lambda)}({\bf r})\,.
\label{schroedinger}
\ee
$E^{(0)}_H$ is the leading-order binding energy of the quarkonium $H$: $M_H = 2m + E^{(0)}_H$.
For later use, we write $\Phi_{H(\lambda)}^{(0)}$ for $L=0$ states,
\bea
\Phi_{n^3S_1(\lambda)}^{(0)}({\bf r})  &=& \frac{1}{\sqrt{4\pi}} \, R_{n0}(r)  \, 
\frac{\bfsigma \cdot {\bf e}_{n^3S_1}(\lambda)}{\sqrt{2}}\,,
\label{wavefun:jpsi}
\\
\Phi_{n^1S_0}^{(0)}({\bf r})  &=& \frac{1}{\sqrt{4\pi}} \, R_{n0}(r)  \, 
\frac{1}{\sqrt{2}}
\label{wavefun:etac}
\,,
\eea
where ${\bf e}_{n^3S_1}(\lambda)$ is the polarization vector of 
the state $n^3S_1$, normalized as ${\bf e}_{n^3S_1}^*(\lambda)\cdot {\bf e}_{n^3S_1}(\lambda^\prime)
= \delta_{\lambda\lambda'}$, and for $L=1$ states, 
\bea
\Phi_{n^1P_1(\lambda)}^{(0)}({\bf r}) &=& 
\sqrt{\frac{3}{4\pi}} \, R_{n1}(r)  \, 
\frac{{\bf e}_{n^1P_1}(\lambda)\cdot{\hat{\bf r}} }{\sqrt{2}} \,,
\label{wavefun:hc}
\\
\Phi_{n{^3P_0}}^{(0)}({\bf r}) &=& 
\sqrt{\frac{1}{4\pi}} \, R_{n1}(r)  \, 
\frac{\bfsigma \cdot {\hat{\bf r}}}{\sqrt{2}} \,,
\label{wavefun:chic0}
\\
\Phi_{n{^3P_1}(\lambda)}^{(0)}({\bf r}) &=& 
\sqrt{\frac{3}{8\pi}} \, R_{n1}(r)  \, 
\frac{\bfsigma \cdot ({\hat{\bf r}} \times {\bf e}_{n^3P_1}(\lambda))}{\sqrt{2}} \,,
\label{wavefun:chic1}
\\
\Phi_{n{^3P_2}(\lambda)}^{(0)}({\bf r}) &=& 
\sqrt{\frac{3}{4\pi}} \, R_{n1}(r)  \, 
\frac{\bfsigma^i  \, h^{ij}_{n^3P_2}(\lambda) \, \hat{\bf r}^j}{\sqrt{2}} \,,
\label{wavefun:chic2}
\eea
where ${\bf e}_{n^{1}P_1}(\lambda)$ and ${\bf e}_{n^{3}P_1}(\lambda)$ 
are polarization vectors satisfying 
${\bf e}_{n^1P_1}^*(\lambda)\cdot {\bf e}_{n^1P_1}(\lambda^\prime)
= {\bf e}_{n^3P_1}^*(\lambda)\cdot {\bf e}_{n^3P_1}(\lambda^\prime) = \delta_{\lambda\lambda'}$,
whereas the polarization of the $n^3P_2$ state is represented by a
symmetric and traceless rank-2 tensor $h_{n^3P_2}^{ij}(\lambda)$ normalized 
according to $h_{n^3P_2}^{ij\,*}(\lambda) \, h_{n^3P_2}^{ji}(\lambda^\prime) =
\delta_{\lambda\lambda'}$.

The state $\ket{H({\bf P},\lambda)}$ may be obtained from 
$\ket{H({\bf P},\lambda)}^{(0)}$ by quantum-mechanical perturbation theory.
At relative order $v^2$ the following corrections may be relevant.

\subsubsection{Higher-order potentials}
In the weak-coupling regime, corrections to the zeroth-order pNRQCD Hamiltonian of the type
$$
\delta H = 
 \int d^3R \int d^3r \, {\rm Tr}\, \left\{
{\rm S}^\dagger \,  \delta V_S \, {\rm S} \right\}\,,
$$
are typically suppressed by  $v^2$ with respect to the leading term.
First-order corrections to the quarkonium state, induced by these terms, are 
therefore relevant at relative order $v^2$:
\be
\ket{H({\bf P},\lambda)}^{(1)} = \int d^3R \int d^3r \; e^{i{\bf P}\cdot {\bf R}} \, {\rm Tr} \left\{
\delta\Phi_{H(\lambda)}({\bf r})\,{\rm S}^\dagger({\bf r},{\bf R})\ket{0}\right\}\,,
\label{ketH1}
\ee
where 
\be
\delta  \Phi_{H(\lambda)}({\bf r}) = 
\sum_{H'\neq H, \lambda'}\frac{\Phi_{H'(\lambda')}^{(0)}({\bf r})}{E_H^{(0)} - E_{H'}^{(0)}}
\; \braQM{H'(\lambda')} \delta V_S \ketQM{H(\lambda)}\,,
\label{QM1}
\ee
and $\braQM{{\bf r}} H(\lambda)\rangle = \Phi^{(0)}_{H(\lambda)}({\bf r})$.
Here and in the following, we shall use the Dirac ket to
indicate the eigenstate either of a {\it quantum-mechanical} operator (like 
$\ket{{\bf r}}$, which stands for an eigenstate of the position operator, 
or $\ketQM{H(\lambda)}$, sometimes also written as $ \ket{nL}$, which stands
for an eigenstate of $ h_S^{(0)}$) or of a {\it quantum-field} operator 
(like $\ket{H({\bf P},\lambda)}$, which stands for an 
eigenstate of the pNRQCD Hamiltonian).

In general, $\delta V_S$ may also depend on the
centre-of-mass momentum ${\bf P}$. We shall distinguish between zero-recoil
corrections (where $\delta V_S$ does not depend on ${\bf P}$) and (final-state)
recoil corrections (where $\delta V_S$ depends on ${\bf P}$).
Effects of these corrections to the transition amplitude will be discussed in Sec.~\ref{wave:potential}.

\subsubsection{Higher Fock-space components}
\label{sec:fock}
The leading correction to the quarkonium state that accounts for the octet component 
is induced by 
\be
\delta H = 
 - \int d^3R \int d^3r \, {\rm Tr}\, \left\{
{\rm O}^\dagger \,  {\bf r}\cdot g{\bf E} \, {\rm S} 
+ {\rm S}^\dagger \,  {\bf r}\cdot g{\bf E} \, {\rm O} 
\right\}\,.
\label{QCDE1}
\ee
According to the power counting, this is a correction of relative order $v$.
The first-order correction to the quarkonium state is
\bea
\hspace{-10mm}
\ket{H({\bf P},\lambda)}^{(1)} &=& \int d^3R \int d^3r  \; e^{i{\bf P}\cdot
  {\bf R}} \, \int d^3x \; {\rm Tr} \, \Bigg\{
{\rm O}^\dagger ({\bf r},{\bf R})
\braQM{{\bf r}} \frac{1}{E_H^{(0)} - h_O^{(0)} - H_{\rm light}}\ketQM{{\bf x}} 
\nn\\
&& \qquad\qquad\qquad \times
\left[-{\bf x}\cdot g{\bf E}({\bf R})\right] \; \Phi^{(0)}_{H(\lambda)}({\bf x})\,\ket{0}\Bigg\}\,.
\label{ketH1fock}
\eea
Since it has a vanishing projection on $\ket{H({\bf P},\lambda)}^{(0)}$, in a transition matrix element 
it contributes at relative order $v^2$. Second-order corrections are of
relative order $v^2$. They contain two orthogonal parts:
\be
\ket{H({\bf P},\lambda)}^{(2)} =
\ket{H({\bf P},\lambda)}^{(2)}_\perp + \ket{H({\bf P},\lambda)}^{(2)}_\parallel\,, 
\label{ketH2fock}
\ee
where
\bea 
&& \hspace{-10mm}
\ket{H({\bf P},\lambda)}^{(2)}_\perp = 
\int d^3R \int d^3r \; e^{i{\bf P}\cdot {\bf R}} \, 
\int d^3x \int d^3y \; {\rm Tr} \, \Bigg\{
{\rm S}^\dagger({\bf r},{\bf R})
\nn\\
&& \qquad\qquad\qquad \times
\braQM{{\bf r}} \sum_{H'\neq H, \lambda'}
\frac{\Phi_{H'(\lambda')}^{(0)}({\bf r})\Phi_{H'(\lambda')}^{(0)\,*}({\bf y})}{E_H^{(0)} - E_{H'}^{(0)} - H_{\rm light}}
\ketQM{{\bf y}} 
\; \left[-{\bf y}\cdot g{\bf E}({\bf R})\right] \; 
\nn\\
&& \qquad\qquad\qquad \times
\braQM{{\bf y}} 
\frac{1}{E_H^{(0)} - h_O^{(0)} - H_{\rm light}}
\ketQM{{\bf x}}
\; \left[-{\bf x}\cdot g{\bf E}({\bf R})\right] \; 
\Phi^{(0)}_{H(\lambda)}({\bf x})\,\ket{0}\Bigg\}\,,
\\
&&\nn
\\
&& \hspace{-10mm}
\ket{H({\bf P},\lambda)}^{(2)}_\parallel =
\frac{\delta Z_{H(\lambda)}}{2}\; \ket{H({\bf P},\lambda)}^{(0)} \,,
\\
&& \hspace{-10mm}
\left( 1+\frac{\delta Z_{H(\lambda)}}{2} \right)\, (2\pi)^3\delta^3({\bf P}-{\bf P}') 
= ~ ^{(0)}\bra{H({\bf P}',\lambda)} H({\bf P},\lambda)\rangle
= \sqrt{\frac{\partial E_{H(\lambda)}}{\partial E_H^{(0)}}}\, (2\pi)^3\delta^3({\bf P}-{\bf P}') 
\,.
\label{ZH}
\eea
$1+ \delta Z_{H(\lambda)}$ is the usual quantum-mechanical normalization
constant of the state, which stands for the probability to find the leading 
color-singlet component in a physical quarkonium state. 
Effects of these corrections to the transition amplitude will be discussed in Sec.~\ref{wave:octet}.

\subsection{M1 transitions in the non-relativistic limit}
\label{M1:NRlimit:pNRQCD}
In accordance with the power counting of pNRQCD, the leading contribution to 
M1 transitions comes from 
\be
{\cal L}^{(0)}_{\rm M1} = 
\int d^3r \;  {\rm Tr} \, \left\{
\frac{1}{2 m}
\; V^{\frac{\sigma\cdot B}{m}}_S 
\; \left\{{\rm S}^\dagger , \bfsigma \cdot e e_Q {\bf B}^{\em}\right\} {\rm S} 
\right\}\,.
\label{M1:lowest}
\ee
As we will discuss in the next section, at leading order $V^{\frac{\sigma\cdot B}{m}}_S =1$.

For $S$-wave quarkonium, substituting Eqs.~(\ref{ketHP00}), (\ref{wavefun:jpsi}), 
(\ref{wavefun:etac}) and (\ref{M1:lowest}) into
Eq.~(\ref{transitionamplitude}) leads to 
\bea
{\cal A}^{(0)}\left[n^3S_1({\bf 0},\lambda) \to n^{\prime\,1}S_0(-{\bf k})
  \gamma({\bf k},\sigma)\right] 
= \delta_{nn^\prime}\,
i ee_Q \frac{{\bf e}_{n^3S_1}(\lambda)\cdot ({\bf k}\times\bfepsilon^*(\sigma))}{m} 
\,,
\label{ampl:S:Lowestorder}
\eea
where we have used that 
\be
\langle \gamma({\bf k},\sigma) | {\bf B}^{\em}({\bf R}) |0\rangle  
=-i \, {\bf k} \times \bfepsilon^*(\sigma)\,e^{-i{\bf k}\cdot{\bf R}} \,. 
\ee
The factor $\delta_{nn^\prime}$ in Eq.~(\ref{ampl:S:Lowestorder}) 
comes from the overlap integral $\displaystyle \int_0^\infty dr\, r^2 \,R_{n\ell}(r)\,R_{n^\prime\ell}(r)$.
Substituting the transition amplitude into Eq.~(\ref{transitionwidth}), we obtain:
\be
\Gamma_{n^3S_1 \to n^{\prime\,1}S_0 \, \gamma} = 
\delta_{nn^\prime}
\frac{4}{3} \alpha e_Q^2 \frac{k_\gamma^3}{m^2}
\left( 1 - \frac{k_\gamma}{M_{n^3S_1}} \right)\,.
\label{NR:S:width}
\ee
The term $- \delta_{nn^\prime}\, k_\gamma/M_{n^3S_1}$ is negligible at order
$k_\gamma^3v^2/m^2$: it vanishes for hindered transitions and is of order 
$v^4$ for allowed ones. Hence, Eq.~(\ref{NR:S:width}) gives back
Eq.~(\ref{M1:NRlimit}) at leading order in the multipole expansion, 
i.e. the well-known formula of the transition width in the non-relativistic limit. 

For $P$-wave quarkonium we obtain:
\bea
{\cal A}^{(0)}\left[n^3P_0({\bf 0}) 
\to n^{\prime\,1}P_1(-{\bf k},\lambda^\prime) \gamma({\bf k},\sigma)\right] &=& 
\delta_{nn^\prime}\,
i ee_Q \frac{{\bf e}^*_{n^{\prime\,1}P_1}(\lambda^\prime)\cdot ({\bf k}\times\bfepsilon^*(\sigma))}{\sqrt{3}m} 
\,,
\label{ampl:P0:Lowestorder}
\\
{\cal A}^{(0)}\left[n^3P_1({\bf 0},\lambda) 
\to n^{\prime\,1}P_1(-{\bf k},\lambda^\prime) \gamma({\bf k},\sigma)\right] &=& 
\nn\\
& & \hspace{-25mm}
\delta_{nn^\prime}\,
i ee_Q \frac{{\bf e}^*_{n^{\prime\,1}P_1}(\lambda^\prime)\cdot \left[{\bf
    e}_{n^{3}P_1}(\lambda)\times ({\bf k}\times \bfepsilon^*(\sigma))\right]}
{\sqrt{2}m} 
\,,
\label{ampl:P1:Lowestorder}
\\
{\cal A}^{(0)}\left[n^3P_2({\bf 0},\lambda) 
\to n^{\prime\,1}P_1(-{\bf k},\lambda^\prime) \gamma({\bf k},\sigma)\right] &=& 
\nn\\
& & \hspace{-25mm}
\delta_{nn^\prime}\,
i ee_Q \frac{{\bf e}^{*\,i}_{n^{\prime\,1}P_1}(\lambda^\prime)\,
h_{n^3P_2}^{ij}(\lambda)\, ({\bf k}\times \bfepsilon^*(\sigma))^j}{m} 
\,.
\label{ampl:P2:Lowestorder}
\eea
Substituting the transition amplitudes into Eq.~(\ref{transitionwidth}), we
end up with 
\be
\Gamma_{n^3P_J \to n^{\prime\,1}P_1 \, \gamma} = 
\delta_{nn^\prime}
\frac{4}{3} \alpha e_Q^2 \frac{1}{m^2}k_\gamma^3
\left( 1 - \frac{k_\gamma}{M_{n^3P_J}} \right)\,.
\label{NR:P:width}
\ee 
The width for $P$-wave spin-singlet to spin-triplet transitions is 
obtained by multiplying the right-hand side of Eq.~(\ref{NR:P:width}) by $(2J+1)/3$.

In the following, we will concentrate on higher-order corrections to $S$-wave transitions.  
We shall come back to $P$-wave transitions in Sec.~\ref{sec:final:P}.

\section{Matching of pNRQCD magnetic dipole operators}
\label{matching:NRQCD:PNRQCD}
Our aim is to complete Eq.~(\ref{NR:S:width}) with corrections of relative
order $v^2$. In accordance with the counting $\als(m) \sim v^2$ and $\als(1/r) \sim v$,
these also include corrections to the matching coefficients of pNRQCD. 

The matching coefficients of pNRQCD encode gluons 
of energy or momentum of order $mv$. Since this scale is associated with the distance between the two 
heavy quarks, the matching coefficients are, in general, functions of $r$.
They also contain hard contributions, typically encoded in the matching coefficients 
inherited from NRQCD. In the following, we will retain the full matching
coefficients of NRQCD and count the matching from NRQCD to pNRQCD only 
in powers of $\als$ calculated at the soft scale. We will exploit 
the explicit form of the NRQCD matching coefficients in Secs.~\ref{sec:final} and \ref{sec:phen}.

\begin{figure}
\makebox[-8cm]{\phantom b}
\put(5,0){\epsfxsize=6truecm\epsffile{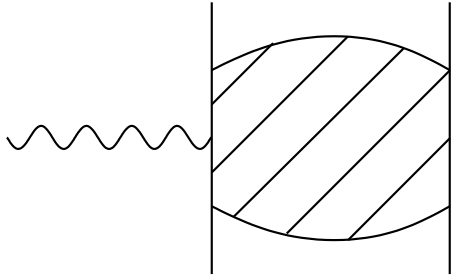}}
\caption{Green function with four external quark/antiquark lines and a photon. 
 The time flows from the bottom to the top.}
\label{figGgamma}
\end{figure}

The matching from NRQCD to pNRQCD may be performed by calculating Green
functions in the two theories and imposing that they are equal order by order
in the inverse of the mass and in the multipole expansion.
In particular, to match the electromagnetic couplings, we need 
Green functions with  four external quark/antiquark lines and one external photon line 
as shown in Fig.~\ref{figGgamma}. The matching condition reads:
\be
G_\gamma^{\rm NRQCD}(x_1',x_2',x_1,x_2) = G_\gamma^{\rm pNRQCD}(x_1',x_2',x_1,x_2)\,.
\ee
Since we are working in a situation where the typical momentum transfer between 
the heavy quarks is larger than $\lQ$, we may, in addition,
perform the matching order by order in $\als$. 

\subsection{Matching at ${\cal O}(1)$}
If we aim at calculating the matching at ${\cal O}(1)$ a convenient  
approach consists in projecting NRQCD on the two-quark Fock space spanned by 
\be
\int d^3x_1\,d^3x_2 \, \psi^{\dagger}(\boldsymbol{x}_1,t)
\varphi(\boldsymbol{x}_1,\boldsymbol{x}_2,t) \chi(\boldsymbol{x}_2,t) \ket{{0}}\,,
\label{FockQQbar}
\ee
where $\varphi(\boldsymbol{x}_1,\boldsymbol{x}_2,t)$ is a $3 \otimes 3$ tensor
in color space and a $2 \otimes 2$ tensor in spin space. 
After projection, all gluon fields are multipole expanded in ${\boldsymbol{r}}$.
Gauge invariance is made explicit at the Lagrangian level by decomposing    
\bea
\varphi(\boldsymbol{x}_1, \boldsymbol{x}_2,t)  
&=& {\rm P}\, \exp \left(ig\int_{x_1}^{x_2} \!\! {\bf A} \cdot d{\bf x} \right) \,{\rm S}^\prime({\bf R},{\bf r},t)
\nn \\
&&
+ {\rm P} \, \exp \left(ig\int_{\bf R}^{x_1} \!\! {\bf A} \cdot d{\bf x} \right) \,{\rm O}^\prime({\bf R},{\bf r},t)\,
\exp\left(ig\int_{x_2}^{\bf R} \!\! {\bf A} \cdot d{\bf x} \right)\!,
\label{Phi:to:S_and_O}
\\
S^\prime({\bf R},{\bf r},t)
&=& \exp\left( iee_Q\int_{x_1}^{x_2} \!\!{\bf A}^{\em} \cdot d{\bf x} \right)\,
S({\bf R},{\bf r},t)  \,,
\label{PhiS:to:S_and_O}
\\
O^\prime({\bf R},{\bf r},t) 
&=& \exp\left( iee_Q\int_{x_1}^{x_2} \!\!{\bf A}^{\em} \cdot d{\bf x} \right)\,
O({\bf R},{\bf r},t)  \,,
\label{PhiO:to:S_and_O}
\eea
where ${\rm P}$ stands for path ordering, ${\bf R} = ({\bf x}_1+{\bf x}_2)/2$ and ${\bf r} = {\bf x}_1-{\bf x}_2$. 
The fields ${\rm S}$ and ${\rm O}$ transform like singlets  under $U(1)_{\em}$ gauge transformations and like singlet 
and octet respectively under $SU(3)_{\rm c}$ gauge transformations.
After projecting (\ref{NRQCD:Lag}) on (\ref{FockQQbar}), one obtains 
\bea
&& V_A^\em = 1\,,
\\
&& V^{\frac{\sigma\cdot B}{m}}_S = c_F^\em\,,
\label{VSB}
\\
&& V^{(r\cdot \nabla)^2 \frac{\sigma\cdot B}{m}}_S =c_F^\em\,,
\\
&& V^{\frac{\sigma\cdot B}{m}}_O = c_F^\em\,,
\\
&& V^{\frac{\sigma\cdot \nabla \times E}{m^2}}_S =  c_S^\em\,,
\label{csem1}
\\
&& 
V^{\frac{\sigma\cdot \nabla_r \times r\cdot \nabla E}{m^2}}_S =  c_S^\em\,,
\label{csem2}
\\
&& V^{\frac{\nabla_r^2 \, \sigma\cdot B}{m^3}}_S = c_{W1}^\em - c_{W2}^\em = 1\,,
\\
&& V^{\frac{(\nabla_r\cdot\sigma)\, (\nabla_r\cdot B)}{m^3}}_S = c^\em_{p^\prime p}\,.
\eea
The matching coefficients  $V^{\frac{\sigma\cdot (r \times r \times B)}{m^2}}_{S}$ and
$V^{\frac{\sigma\cdot B}{m^2}}_{S}$ are zero at ${\cal O}(1)$.

We consider now the impact of the ${\cal O}(1)$ matching on the transition amplitude.
In order to keep the notation compact, it is useful to define
\be
\overline{\cal A} \equiv \frac{\cal A}{\bar{\cal A}^{(0)}}\,,
\ee
where ${\cal A}$ is an amplitude calculated from Eq.~(\ref{transitionamplitude}) by substituting 
${\cal L}_{\gamma\,\rm pNRQCD}$ with the considered operator and ${\bar{\cal
    A}^{(0)}}$ is connected with the leading M1 amplitude ${\cal A}^{(0)}$
(\ref{ampl:S:Lowestorder}) by 
\be
\delta_{nn'}\bar{\cal A}^{(0)} \equiv {\cal A}^{(0)}\,.
\label{A0bar}
\ee
\begin{itemize}
\item[(A)]{ 
The matching coefficient (\ref{VSB}) induces the following correction 
to ${\cal A}^{(0)}$:
\bea
\hspace{-5mm}
\overline{\cal A}\left[n^3S_1({\bf 0},\lambda) 
\to n^{\prime\,1}S_0(-{\bf k}) \gamma({\bf k},\sigma)\right]
 &=& 
\kappa_Q^\em \, \delta_{nn'}
\,.
\label{ampl:S:1}
\eea
}
\item[(B)]{ 
The correction induced by the operator $\displaystyle \frac{c_F^\em}{16 m}
\; \left\{ {\rm S}^\dagger , {\bf r}^i {\bf r}^j 
    (\bfnabla^i \bfnabla^j \bfsigma \cdot e e_Q {\bf B}^{\em})\right\} {\rm S}$ is 
\bea
\hspace{-5mm}
\overline{\cal A}\left[n^3S_1({\bf 0},\lambda) 
\to n^{\prime\,1}S_0(-{\bf k}) \gamma({\bf k},\sigma)\right]
 &=& 
-\frac{c_F^\em}{24}\,k^2\, \braQM{n'S} {\bf r}^2 \ketQM{nS}
\,.
\label{ampl:S:2}
\eea
}
\item[(C)]{ 
The correction induced by the operator 
$\displaystyle - \frac{c_S^\em}{16 m^2} 
\; \left[{\rm S}^\dagger,  \bfsigma \cdot \left[-i\bfnabla \times, e e_Q {\bf E}^\em \right]\right] {\rm S}$ is 
\bea
\hspace{-5mm}
\overline{\cal A}\left[n^3S_1({\bf 0},\lambda) 
\to n^{\prime\,1}S_0(-{\bf k}) \gamma({\bf k},\sigma)\right]
 &=& 
c_S^\em\, \frac{k}{8m}\,\delta_{n'n}
\,,
\label{ampl:S:3}
\eea
where we have used 
\be
\langle \gamma({\bf k},\sigma) | {\bf E}^{\em}({\bf R}) |0\rangle  
=-i k \bfepsilon^*(\sigma)\,e^{-i{\bf k}\cdot{\bf R}} \,. 
\ee
}
\item[(D)]{ 
The correction induced by the operator 
$\displaystyle - \frac{c_S^\em}{16 m^2} 
\; \left[ {\rm S}^\dagger,  \bfsigma \cdot \left[-i\bfnabla_r \times, 
{\bf r}^i (\bfnabla^i e e_Q  {\bf E}^\em) \right]\right] {\rm S}$ is 
\bea
\hspace{-7mm}
\overline{\cal A}\left[n^3S_1({\bf 0},\lambda) 
\to n^{\prime\,1}S_0(-{\bf k}) \gamma({\bf k},\sigma)\right]
 &=& 
c_S^\em\,\frac{k}{8m}\,\left( \delta_{n'n} + i \frac{2}{3} \, \braQM{n'S} {\bf r}\cdot{\bf p}  \ketQM{nS} \right).
\label{ampl:S:4}
\eea
}
\item[(E)]{ 
The correction induced by the operator 
$\displaystyle \frac{1}{4 m^3} \; \left\{ {\rm S}^\dagger , \bfsigma \cdot e e_Q {\bf
  B}^{\em} \right\} \bfnabla_r^2 {\rm S}$ is
\bea
\hspace{-5mm}
\overline{\cal A}\left[n^3S_1({\bf 0},\lambda) 
\to n^{\prime\,1}S_0(-{\bf k}) \gamma({\bf k},\sigma)\right]
 &=& 
- \braQM{ n'S} \frac{{\bf p}^2}{2m^2}  \ketQM{nS}
\,.
\label{ampl:S:5}
\eea
}
\item[(F)]{ 
The correction induced by the operator 
$\displaystyle \frac{c_{p^\prime p}^\em}{8 m^3} 
\; \left\{ {\rm S}^\dagger , \bfsigma^i \, e e_Q {\bf
  B}^{\em\,j} \right\} \bfnabla_r^i\bfnabla_r^j {\rm S}$ is
\bea
\hspace{-5mm}
\overline{\cal A}\left[n^3S_1({\bf 0},\lambda) 
\to n^{\prime\,1}S_0(-{\bf k}) \gamma({\bf k},\sigma)\right]
 &=& 
- \frac{\kappa^\em_Q}{3} \braQM{n'S} \frac{{\bf p}^2}{2m^2}  \ketQM{nS}
\,,
\label{ampl:S:6}
\eea
where we have used $c_{p'p}^\em = \kappa^\em_Q$.
}
\end{itemize}

\subsection{Calculation of $V^{\frac{\sigma\cdot B}{m}}_S$}
\label{sec:VSB}
In this section, we match the operator (\ref{M1:lowest}) beyond ${\cal O}(1)$.
This operator provides the leading transition widths (\ref{NR:S:width}) and (\ref{NR:P:width}) 
in the case of allowed M1 transitions. Hence, corrections of order $\als$ and $\als^2$ to 
$V^{\frac{\sigma\cdot B}{m}}_S$, which may arise in the matching from NRQCD to
pNRQCD, are potentially larger than or of the same order as genuine
relativistic $v^2$ corrections to the transition width. 
Surprisingly, we shall be able to perform the matching exactly and provide 
a result that is valid to all orders in perturbation theory and non-perturbatively.

Before going to the matching, we recall that the 
matching coefficient $c_F^\em$ that appears in the NRQCD Lagrangian
(\ref{NRQCD:Lag}) is the heavy-quark magnetic moment.
The matching coefficient $V^{\frac{\sigma\cdot B}{m}}_S$
is the magnetic moment of the singlet-quarkonium field.
While the first one gets only contributions from the hard modes,
the second one may potentially get large contributions steming from the soft scale.

The matching of $V^{\frac{\sigma\cdot B}{m}}_S$ proceeds as follows.

{\bf (1)} First, we note that the only amplitudes in NRQCD that 
may contribute to the matching are those where the photon couples to 
the heavy quark (antiquark) through the operators $\psi^\dagger \bfsigma \cdot e e_Q {\bf B}^{\em} \psi /m$ 
or $-\chi^\dagger \bfsigma \cdot e e_Q {\bf B}^{\em} \chi /m$. At order $1/m$, 
this is the only magnetic spin-flipping coupling to the quark. 
If the photon couples to loops of massless quarks then, 
at leading order in the electromagnetic coupling constant,
the sum of the electric charges over three light flavors vanishes. In the bottomonium system, 
contributions from charm-quark loops should also be considered. If the  
momentum flowing through the loop is hard, the contribution is suppressed by,
at least,  $\als^2(m_b)$ and, therefore, beyond the accuracy of this work. If the momentum is soft, 
the charm quark effectively decouples and the system can be described by an
effective field theory with three massless quarks \cite{Beneke:1999zr,Brambilla:2001qk}.

\begin{figure}
\makebox[-14cm]{\phantom b}
\put(5,0){\epsfxsize=13truecm\epsffile{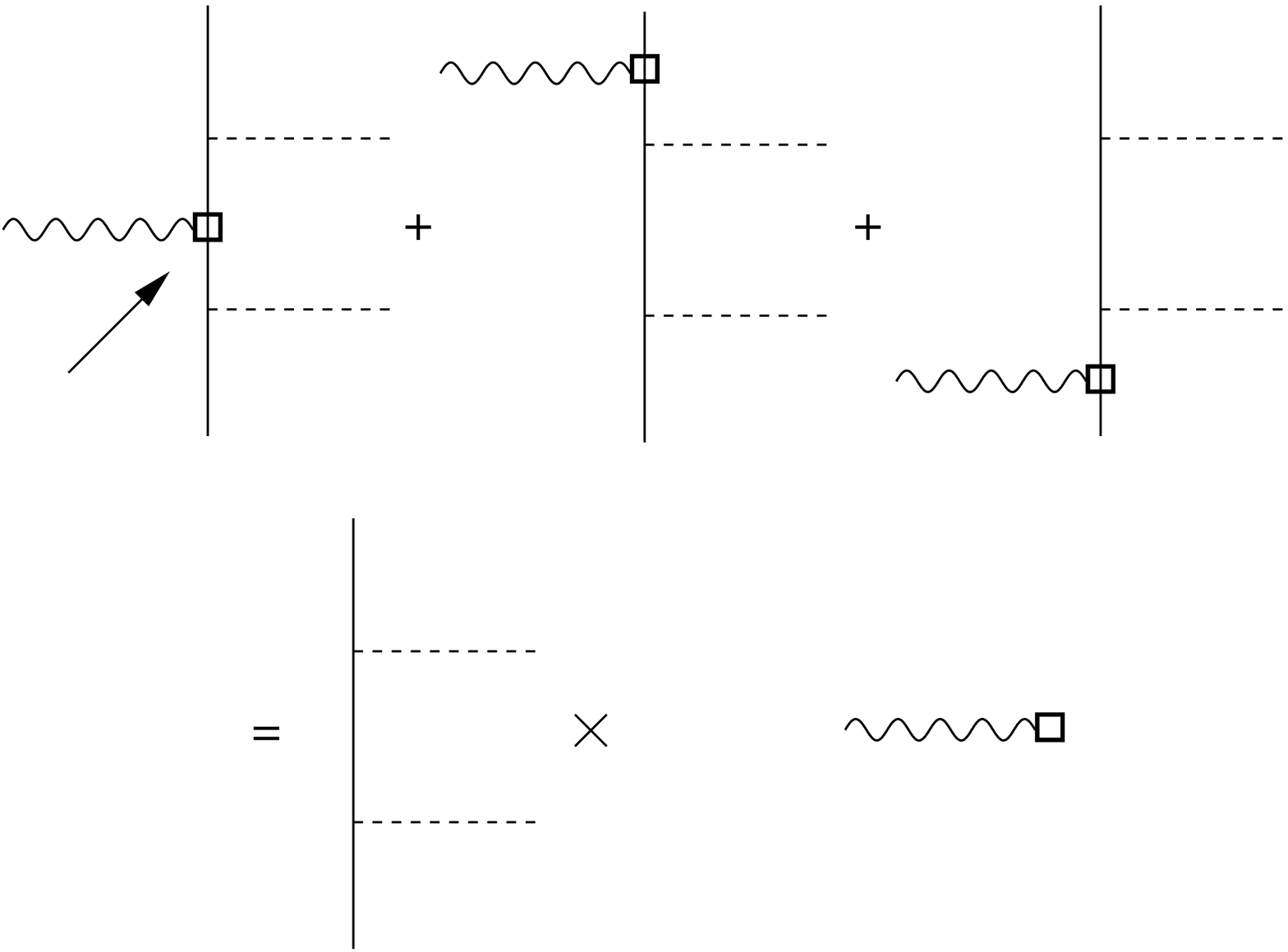}}
\put(-20,140){$\displaystyle c_F^\em \frac{\bfsigma\cdot ee_Q{\bf B}^\em}{2m}$}
\put(72,268){$t_f$}
\put(72,238){$t_1$}
\put(72,203){$t$}
\put(72,173){$t_2$}
\put(72,143){$t_i$}
\put(-35,203){$\displaystyle \int_{t_i}^{t_f} \!\!dt$}
\put(205,62){$\displaystyle \int_{t_i}^{t_f} \!\! dt$}
\caption{Diagrammatic factorization of the magnetic dipole coupling in static
  NRQCD: $t_f > t_1 > t_2 >t_i$. Dashed lines are longitudinal gluons, the box represents the
  magnetic dipole coupling.}
\label{figfac}
\end{figure}

{\bf (2)} The crucial point is to recognize that there are no extra 
momentum or spin dependent insertions on the heavy-quark lines that 
contribute to the matching of $V^{\frac{\sigma\cdot
    B}{m}}_S$, since they carry extra $1/m$ suppressions. As a consequence, 
the magnetic spin-flipping operator $\bfsigma \cdot e e_Q {\bf B}^{\em}/m$ 
behaves, for the purpose of the matching, as the identity operator in
coordinate space\footnote{
The property that $\bfsigma \cdot e e_Q {\bf B}^{\em}/m$ depends neither 
on gluon fields nor on the relative coordinate ${\bf r}$, which in turn is a
consequence of the ultrasoft nature of the external photon, will be used again and again 
in the course of the paper and is responsible for most of results.} 
and the magnetic matrix element factorizes. 
Therefore, any (normalized) NRQCD amplitude contributing 
to $V^{\frac{\sigma\cdot B}{m}}_S$ may be written as 
\be 
\frac{c_F^\em}{2 m}  \int^{t_f}_{t_i} dt \,\langle \gamma| \bfsigma^{(1)} \cdot e e_Q {\bf B}^{\em}({\bf x}_1,t)
+ \bfsigma^{(2)} \cdot e e_Q {\bf B}^{\em}({\bf x}_2,t) |0\rangle \,,
\label{SBmatch1}
\ee
which, after multipole expansion of the magnetic field, becomes (we
neglect terms proportional to ${\bf r}$):
\be
\left(\frac{c_F^\em}{2m} + \frac{c_F^\em}{16m}
\left({\bf r}\cdot\bfnabla\right)^2 + \dots \right) 
\left(\bfsigma^{(1)} + \bfsigma^{(2)}\right) \cdot 
\int^{t_f}_{t_i} dt\,
\langle \gamma| e e_Q {\bf B}^{\em}(t)|0\rangle \,,
\label{sb2}
\ee
where $\bfsigma^{(1)}$ stands for the Pauli matrices acting on the quark and 
$\bfsigma^{(2)}$ for the Pauli matrices acting on the antiquark. 

To see how factorization works from a diagrammatic point of view, 
let us consider a photon insertion on a quark line. In general, this happens in between
two longitudinal gluon insertions (first diagram of Fig.~\ref{figfac}). 
Transverse gluons couple to quark lines through $1/m$ suppressed operators
and are irrelevant for the purpose of the matching. At order $1/m^0$,  
the coupling of longitudinal gluons to quark lines is spin and momentum
independent. Therefore, $\bfsigma \cdot e e_Q {\bf B}^{\em}$ may be freely moved along
the quark lines. The sum of the first three diagrams of Fig.~\ref{figfac} is proportional to
\bea
&& 
\theta(t_f-t_1) \theta(t_1-t) \theta(t-t_2) \theta(t_2-t_i) +  
\theta(t_f-t) \theta(t-t_1) \theta(t_1-t_2) \theta(t_2-t_i) 
\nn\\
&& \qquad 
+  \theta(t_f-t_1) \theta(t_1-t_2) \theta(t_2-t) \theta(t-t_i)  
\nn\\
&& =   
\theta(t_f-t) \theta(t-t_i) \theta(t_f-t_1) \theta(t_1-t_2) \theta(t_2-t_i) \,,
\nn
\eea
where the thetas come from the static heavy-quark propagators.
The equality completes the factorization proof graphically represented in  Fig.~\ref{figfac}. 

{\bf (3)} By matching Eq.~(\ref{sb2}) to the pNRQCD amplitude 
\be
\left(\frac{V^{\frac{\sigma\cdot B}{m}}_S}{2m} + 
\frac{V^{(r\cdot \nabla)^2 \frac{\sigma\cdot B}{m}}_S}{16m}
\left({\bf r}\cdot\bfnabla\right)^2 + \dots \right)
\left(\bfsigma^{(1)} + \bfsigma^{(2)}\right) \cdot 
\int^{t_f}_{t_i} dt \,\langle \gamma| e e_Q {\bf B}^{\em}(t)|0\rangle \,,
\label{sb1}
\ee
it follows that 
\be
V^{\frac{\sigma\cdot B}{m}}_S = V^{(r\cdot \nabla)^2 \frac{\sigma\cdot
    B}{m}}_S  = c_F^\em\,.
\label{m11}
\ee

Equation (\ref{m11}) is a result that holds to all orders in the strong-coupling
constant and also non-perturbatively. It excludes that the $1/m$ magnetic coupling 
of the quarkonium field is affected by any soft contribution.
A fortiori, it excludes large anomalous non-perturbative corrections to this coupling.

\begin{figure}
\makebox[-15cm]{\phantom b}
\put(0,5){\epsfxsize=15truecm \epsffile{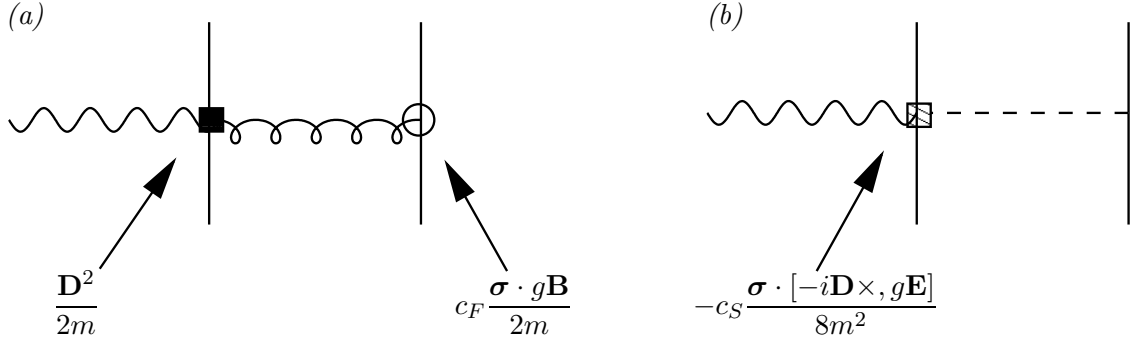}}
\put(18,-10){$\displaystyle \frac{{\bf D}^2}{2m}$}
\put(170,-10){$\displaystyle c_F \frac{\bfsigma\cdot g{\bf B}}{2m}$}
\put(260,-10){$\displaystyle - c_S \frac{\bfsigma\cdot[-i{\bf D}\times, g{\bf E}]}{8m^2}$}
\put(0,100){\it (a)}
\put(265,100){\it (b)}
\caption{Diagrams contributing at order $\als$ to $V^{\frac{\sigma\cdot (r \times r \times B)}{m^2}}_{S}$.
Although not displayed, the symmetric diagrams are understood.
{\bf D} is the covariant derivative under $SU(3)_{\rm c}\times U(1)_{\em}$.}
\label{spin:orbit:matching}   
\end{figure}

\subsection{Calculation of $V^{\frac{\sigma\cdot (r \times r \times B)}{m^2}}_{S}$ and
$V^{\frac{\sigma\cdot B}{m^2}}_{S}$}
\label{M1:assoc:LS:potential}
The matching coefficients $V^{\frac{\sigma\cdot (r \times r \times B)}{m^2}}_{S}$ and
$V^{\frac{\sigma\cdot B}{m^2}}_{S}$ do not get contributions at ${\cal O}(1)$.
At order $\als$, the diagrams contributing to the NRQCD part of the matching are shown 
in Fig~\ref{spin:orbit:matching}. Note that the photon is emitted by the electromagnetic 
field embedded in the covariant derivative. If we sandwich the diagrams between initial 
and final states that are gauge invariant under $SU(3)_{\rm c}\times U(1)_{\em}$
and multipole expand the external electromagnetic field, we obtain:
\bea
&& V^{\frac{\sigma\cdot (r \times r \times B)}{m^2}}_{S} 
= \frac{C_F \als}{2} \, (2c_F-c_S) = \frac{C_F\als}{2}\,,
\label{VLSa}
\\
&& V^{\frac{\sigma\cdot B}{m^2}}_{S} = 0\,.
\label{VLSb}
\eea
In the first equation, we have made use of Eq.~(\ref{rpi:cs}).
Alternatively, we may first perform the matching in a non explicitly gauge-invariant fashion, 
as it is customary in perturbative calculations of Green functions, and then 
impose gauge invariance at the level of the pNRQCD Lagrangian through field 
redefinitions of the type (\ref{Phi:to:S_and_O})-(\ref{PhiO:to:S_and_O}).

Equation (\ref{VLSa}) may be generalized to all orders as
\be
V^{\frac{\sigma\cdot (r \times r \times B)}{m^2}}_{S} 
= \frac{r^2V_S^{(0)\,\prime}}{2}\,,
\label{VLSall}
\ee
where $V_S^{(0)\,\prime}$ stands for $dV_S^{(0)}/dr$.
Also Eq.~(\ref{VLSb}) is valid to all orders. The proof proceeds as follows.

{\bf (1)} The matching can be performed order by order in $1/m$.
NRQCD amplitudes that may contribute to the matching involve either 
insertions of two of the operators ${\bf D}^2/2m$, $c_F \, \bfsigma \cdot g {\bf B}/2m$ and 
$c_F^\em \, \bfsigma \cdot ee_Q {\bf B}^\em/2m$, or one insertion of the operator 
$-c_s\bfsigma\cdot[-i{\bf D}\times,g{\bf E}]/8m^2$ on either the quark or the antiquark line.
Couplings of the photon to massless quark loops or charm-quark loops may be neglected by the same
arguments given in the previous section for the matching of $V^{\frac{\sigma\cdot B}{m}}_S$.

{\bf (2)} First, we consider amplitudes with an insertion of the operator 
$c_F^\em \, \bfsigma \cdot ee_Q {\bf B}^\em/2m$ and one of either 
${\bf D}^2/2m$ or $c_F \, \bfsigma \cdot g {\bf B}/2m$.
In the first case, due to the ultrasoft nature of the external photon, we may 
neglect the action of $\bfnabla$ on ${\bf B}^\em$. Hence the magnetic 
dipole operator behaves like the identity operator, and, following 
an argument similar to that one developed in the previous section, we can show that 
there is no contribution to the matching. In the second case, the QCD part of the amplitude 
factorizes in a term proportional to the expectation value of the chromomagnetic operator, 
which vanishes for parity.

{\bf (3)} In all the remaining terms, the electromagnetic coupling is embedded
in a covariant derivative. Since by projecting the operator
\be
\int d^3x \; \psi^\dagger \left( -i\bfnabla -ee_Q{\bf A}^\em \right) \psi 
+ [\psi \rightarrow i \sigma ^2 \chi^*, {\bf A}^\em \rightarrow - {\bf A}^{\em}]\,,
\label{NRQCDLS}
\ee
onto the state (\ref{FockQQbar}) and taking into account gauge invariance 
(see Eq.~(\ref{PhiS:to:S_and_O}))  we obtain\footnote{
From 
$$
\exp\left( -iee_Q\int_{x_1}^{x_2} {\bf A}^\em \cdot d{\bf x}\right)
\left[ -i\bfnabla_{x_1} - ee_Q{\bf A}(x_1)-i\bfnabla_{x_2} + ee_Q{\bf A}(x_2) \right] 
\exp\left( iee_Q\int_{x_1}^{x_2} {\bf A}^\em \cdot d{\bf x}\right)
$$
and multipole expanding one ends up with 
$
-i\bfnabla + {\bf r}\times ee_Q {\bf B}^\em + {\cal O}(r^2)\,.
$
}  
$$
\int d^3R \, d^3r \; {\rm Tr}\left\{
{\rm S}^\dagger \left( -i\bfnabla + {\bf r}\times ee_Q{\bf B}^\em + {\cal O}(r^2)
\right){\rm S} \right\} \,,
$$
we conclude that the matching coefficients of the operators 
\be
\frac{1}{4 \,m^2\,r^3} \left\{{\rm S}^\dagger , \bfsigma\cdot\left[ {\bf r}
\times  \left({\bf r} \times e e_Q {\bf B}^{\em} \right) \right] \right\} {\rm S} 
\label{m2AA}
\ee
and 
\be
\frac{1}{4 \,m^2\,r^3}
\; \left\{{\rm S}^\dagger , \bfsigma \right\} \cdot\left[ {\bf r} \times  \left(-i \bfnabla\right)
\right] {\rm S} 
\label{m2BB}
\ee
are equal. Indeed, the first is obtained from the second by replacing 
$-i\bfnabla$ by ${\bf r}\times ee_Q{\bf B}^\em$.
The operator $\displaystyle  \frac{1}{4 \,m^2\,r^3}$
$\displaystyle \; \left\{{\rm S}^\dagger , \bfsigma \right\} \cdot\left[ {\bf r} \times  \left(-i \bfnabla\right)
\right] {\rm S} $ is protected by Poincar\'e invariance (Gromes relation) 
\cite{Gromes:1984ma,Brambilla:2001xk,Brambilla:2003nt}. Its matching coefficient is  
equal to  $r^2V_S^{(0)\,\prime}/2$ to all orders in perturbation theory and non-perturbatively.
This proves Eq.~(\ref{VLSall}). It also proves Eq.~(\ref{VLSb}), because the analysis 
does not reveal any contribution to $V^{\frac{\sigma\cdot B}{m^2}}_{S}$.

Equations (\ref{VLSall}) and (\ref{VLSb}) are both valid to all orders 
in perturbation theory and non-perturbatively. The first equation confirms 
earlier findings in phenomenological models (see, for instance, \cite{Grotch:1982bi}).
The second one states that to all orders in the strong-coupling constant and non-perturbatively
the existence of a magnetic coupling of the type induced by a scalar
interaction is excluded. Phenomenological models often assume 
that the relativistic Hamiltonian contains a scalar interaction $\gamma^{0\,(1)}V^{\rm scalar}(r) \gamma^{0\,(2)}$.
The non-relativistic reduction of this term generates, among others, a magnetic spin-flipping term of the type 
$ \displaystyle \frac{1}{2 m^2} \; V^{\rm scalar} 
\left\{{\rm S}^\dagger , \bfsigma \cdot e e_Q {\bf B}^{\em}\right\} {\rm S}$.
Our analysis shows that such a term is excluded from pNRQCD. 
We conclude that a scalar interaction would induce a M1 coupling that in QCD 
cannot be generated, even dynamically, for heavy-quark bound state
systems.\footnote{The situation is here different from the
case of the spin-dependent potentials. There, a spin-orbit potential 
of the type induced by a scalar interaction may be dynamically generated
\cite{Eichten:1980mw,Pineda:2000sz}. 
The particular nature of the scalar interaction contribution to M1 transitions
has also been discussed in \cite{Zhang:1991et}.
}

The correction induced by the operator 
\be
\frac{1}{4 m^2}
\; \frac{r V_S^{(0)\,\prime}(r)}{2} 
\; \left\{{\rm S}^\dagger , \bfsigma\cdot\left[ \hat{\bf r} \times  \left(
  \hat{\bf r}\times e e_Q {\bf B}^{\em} \right) \right] \right\} {\rm S} 
\label{Vm2LS}
\ee
to the M1 transition amplitude is
\bea
\hspace{-5mm}
\overline{\cal A}\left[n^3S_1({\bf 0},\lambda) 
\to n^{\prime\,1}S_0(-{\bf k}) \gamma({\bf k},\sigma)\right]
 &=& 
- \frac{1}{6m} \braQM{n'S} r  V_S^{(0)\,\prime} \ketQM{nS} 
\,.
\label{ampl:S:7}
\eea

Finally, we note that amplitudes with one insertion of the operator  
$- c_S^\em \bfsigma \cdot [-i{\bfnabla}$ $\times, e e_Q {\bf E}^{\em}] /8 m^2$
only contribute to the matching of $V^{\frac{\sigma\cdot \nabla \times E}{m^2}}_S$ 
and $V^{\frac{\sigma\cdot \nabla_r \times r\cdot \nabla E}{m^2}}_S$. Since the
operator factorizes, there are no soft contributions to the matching
coefficients and Eqs.~(\ref{csem1}) and (\ref{csem2}) turn out to be valid to all orders in $\als$.

\subsection{Comment on the matching in the strong-coupling regime}
\label{secstrong}
In the weak-coupling regime, at relative order $v^2$, the only $1/m^3$ operator relevant 
for M1 transitions is\footnote{
The matching coefficients of the operators $1/(4 m^3) \; \{ {\rm S}^\dagger , \bfsigma^i \, e e_Q {\bf
  B}^{\em\,j} \} \bfnabla_r^i\bfnabla_r^j {\rm S}$ and  $1/(m^3 \,r^2) 
\{{\rm S}^\dagger , \bfsigma \cdot e e_Q {\bf B}^{\em}\} {\rm S}$ are suppressed 
by powers of $\als$. Therefore, these operators contribute at relative order $\als\,v^2$ or smaller.
}
$$
\frac{
1}{4 m^3} \; \left\{ {\rm S}^\dagger , \bfsigma \cdot e e_Q {\bf
  B}^{\em} \right\} \bfnabla_r^2 {\rm S} \,.
$$
Note that corrections to the matching coefficient are suppressed by powers of $\als(1/r) \ll 1$.

Comparing our expression of the pNRQCD Lagrangian (\ref{gammapNRQCD:Lag}) with
the phenomenological Hamiltonian used in \cite{Grotch:1982bi}, we observe that, 
up to the scalar interaction term, the two expressions are equal.
The absence of a scalar interaction in pNRQCD has been discussed above.
Here, we remark that our expression is valid in the weak-coupling regime
($mv^2\simg \lQ$) only, while phenomenological Hamiltonians are supposed 
to be applicable to both weakly and strongly coupled quarkonia. We may ask how the pNRQCD 
Lagrangian would change in the strong-coupling regime ($mv \sim \lQ$). 
This has been discussed in the absence of an electromagnetic interaction in 
\cite{Brambilla:2000gk,Pineda:2000sz,Brambilla:2003mu,Brambilla:2001xy,Brambilla:2002nu}. 
Here, we focus on the magnetic dipole couplings. We have shown that the $1/m$ and $1/m^2$
matching is valid beyond perturbation theory. However, this is unlikely to
happen at order $1/m^3$. Since $\als(1/r)\sim 1$ is no longer a suppression factor, 
we expect more NRQCD amplitudes to contribute to the matching.
Among them, we may have amplitudes made of two insertions of the operator
$c_F \, \bfsigma \cdot g {\bf B}/2m$ and one of $c_F^\em \, \bfsigma \cdot
ee_Q {\bf B}^\em/2m$, or one of $-c_s\bfsigma\cdot[-i{\bf D}\times,g{\bf
    E}]/8m^2$ and one of ${\bf D}^2/2m$, or one of $-c_s\bfsigma\cdot[-i{\bf D}\times,g{\bf
    E}]/8m^2$ and one of $c_F \, \bfsigma \cdot g {\bf B}/2m$ and so
on. These amplitudes will be encoded in the matching coefficients of
pNRQCD in the form of static Wilson loop amplitudes with  
field strength insertions of the same kind as those that appear in the QCD potential 
at order $1/m^2$ \cite{Pineda:2000sz}. Also, they may induce new operators in
pNRQCD. A non-perturbative derivation of the pNRQCD Lagrangian coupled 
to the electromagnetic field at order $1/m^3$ has not been worked out.
In a purely analytical approach, such a computation will likely have a limited 
phenomenological impact, due to the many non-perturbative parameters 
(Wilson-loop amplitudes) needed. However, if supplemented by lattice
simulations, it will pave the way for a rigorous QCD study
of relativistic corrections to M1 transitions in excited heavy-quarkonium states.

In summary, phenomenological models used so far to describe magnetic dipole
transitions in quarkonium, once cleaned of the scalar interaction, appear to
be valid only for weakly-coupled resonances. For strongly-coupled resonances, at order 
$1/m^3$, more terms and matching coefficients with, in principle, large 
non-perturbative corrections are expected.

\section{Wave-function corrections to magnetic dipole transitions}
\label{Ov2:RelCorr}
Corrections to the wave function that give contributions of relative order 
$v^2$ to the transition amplitude are of two categories: 
{(A)} higher-order potential corrections, which may be further 
distinguished in {(A.1)} zero-recoil corrections and 
{(A.2)} recoil effects of the final-state quarkonium, 
and {(B)} higher Fock-state corrections.

\subsection{Corrections to the wave function from higher-order potentials}
\label{wave:potential}

\subsubsection{Zero-recoil effects}
We first consider corrections coming from higher-order potentials that do not 
depend on the centre-of-mass momentum of the recoiling quarkonium.
Since $\delta V_S$ is, at least, of order $v^4$, we only need to take into account the correction
induced by Eq.~(\ref{QM1}) to the leading amplitude computed in Sec.~\ref{M1:NRlimit:pNRQCD}. 
The correction is proportional to
$$
(1-\delta_{n'n})\, \braQM{n'S}
{\rm Tr}\,\big\{ \{\bfsigma, \delta V_S  \, {\bf e}_{n^3S_1}(\lambda)\cdot \bfsigma\}  
- \delta V_S \, \{\bfsigma, {\bf e}_{n^3S_1}(\lambda)\cdot \bfsigma\}  \big\} 
\ketQM{nS}\,.
$$
It vanishes for allowed magnetic transitions.
This follows from the fact that $\bfsigma \cdot e e_Q {\bf B}^{\em}/m$ 
is independent of ${\bf r}$ and the first-order correction is orthogonal 
to the zeroth-order wave function (see Eq.~(\ref{QM1})).
The two terms in the trace come from the correction 
to the incoming and outcoming heavy quarkonium respectively.
We distinguish different cases. 
{\it (i)} If $\delta V_S$ is spin independent, then the trace vanishes.
{\it (ii)} If  $\delta V_S$ is a spin-orbit potential, then 
$\braQM{n'S}\delta V_S\ketQM{nS} \sim \braQM{n'S} {\bf r} \times {\bf p}\ketQM{nS} = 0$ on $S$ waves.
{\it (iii)} If  $\delta V_S$ is a tensor potential, then 
$\braQM{n'S}\delta V_S\ketQM{nS} \sim \braQM{n'S} 3\,{\bf r}^i {\bf r}^j - r^2\delta^{ij}
\ketQM{nS} = 0$ on $S$ waves.
{\it (iv)} The only non-vanishing contribution comes from the spin-spin potential:
\be
{\rm Tr} \left\{ {\rm S}^\dagger \, \delta V_S \, {\rm S} \right\} = 
- \frac{V^{\rm ss}({\bf r})}{4\,m^2}  \;
{\rm Tr} \left\{ {\rm S}^\dagger \, \bfsigma^i \, {\rm S} \, \bfsigma^i \right\} \,.
\label{spinspin}
\ee
It induces the following correction to the transition amplitude:
\bea
\hspace{-5mm}
\overline{\cal A}\left[n^3S_1({\bf 0},\lambda) 
\to n^{\prime\,1}S_0(-{\bf k}) \gamma({\bf k},\sigma)\right]
 &=& 
c_F^\em (1-\delta_{n'n}) \frac{1}{m^2} \frac{\braQM{n'S} V^{\rm ss}({\bf
    r})\ketQM{nS}}{E_n^{(0)}-E_{n'}^{(0)}}
\,.
\label{ampl:S:8}
\eea
The correction is only relevant for hindered M1 transitions and, in this case, 
is of order $v^2$ ($V^{\rm ss}/(E_n^{(0)}-E_{n'}^{(0)}) \sim mv^4/mv^2 \sim v^2$).

\subsubsection{Final-state recoil effects}
\label{wave:recoil}
The final-state quarkonium is not at rest. It moves with a velocity $-{\bf k}$ with 
respect to the centre-of-mass frame. In \cite{Grotch:1982bi}, it has been pointed out 
that due to this motion, higher-order potentials that depend on the
centre-of-mass momentum may modify the wave function of the recoiling
quarkonium such that the E1 operator may induce an effective M1 transition.
The leading potential relevant to our case is:
\be
{\rm Tr} \left\{ {\rm S}^\dagger \, \delta V_S \, {\rm S} \right\} = 
- \frac{1}{4 \,m^2} \frac{V_S^{(0)\,\prime}}{2}
\; 
{\rm Tr} \left\{ \{ {\rm S}^\dagger, \bfsigma \}
\cdot\left[ \hat{\bf r} \times  \left(-i \bfnabla\right)
\right] 
{\rm S} \right\}
\,.
\label{LSpotential}
\ee
We have discussed the spin-orbit potential in
Sec.~\ref{M1:assoc:LS:potential} (see Eq.~(\ref{m2BB})), where we noticed that its value is protected 
by Poincar\'e invariance. Inserting Eq.~(\ref{LSpotential}) into 
Eqs.~(\ref{ketH1}) and (\ref{QM1}) we obtain 
\be
\ket{H({\bf P},\lambda)}^{(1)} = 
- \int d^3R \int d^3r \; e^{i{\bf P}\cdot{\bf R}}\;
{\rm Tr} \left\{ \frac{{\bf P}}{8m^2}\cdot 
\left\{ {\rm S}^\dagger, \bfsigma \right\}\times (\bfnabla_r  \Phi_{H(\lambda)})
\ket{0}\right\}\,,
\label{ketHPspin}
\ee
where we have used
$$
\frac{1}{E_H^{(0)} - E_{H'}^{(0)}}
\braQM{H'(\lambda')} \hat{\bf r} V_S^{(0)\,\prime} \ketQM{H(\lambda)} 
= i \, \braQM{H'(\lambda')} {\bf p} \ketQM{H(\lambda)} \,,
$$
which follows from $[{\bf p},h_S^{(0)}] = -i \hat{\bf r}
V_S^{(0)\,\prime}(r)$, and 
$$
\sum_{H'\neq H, \lambda'}\Phi_{H'(\lambda')}^{(0)}({\bf r})\;
i \,\braQM{H'(\lambda')} {\bf p} \ketQM{H(\lambda)} 
= \bfnabla_{r}\Phi_{H(\lambda)}^{(0)}({\bf r})\,,
$$
which follows from completeness and the definite parity of the functions $\Phi_{H(\lambda)}^{(0)}$.
Two different derivations of Eq.~(\ref{ketHPspin}), one that uses Lorentz-boost
transformations and another one based on relativistically covariant formulations, can be found in 
Appendix \ref{AppB}.

Equation (\ref{ketHPspin}) states that, due to the recoil, the final state
develops a nonzero $P$-wave, spin-flipped  component suppressed by a factor $v\,k_\gamma/m$.
As a consequence, in a $n^3S_1 \to n^{\prime\,1}S_0 \, \gamma$ transition, 
the $P$-wave spin-triplet final state component can be reached from the
initial $^3S_1$ state through an E1 transition, mediated by the operator 
\be
{\cal L}_{\rm E1}^{(0)} =  \int d^3 r \;  {\rm Tr} \, \left\{ 
{\rm S}^\dagger {\bf r}\cdot e e_Q {\bf E}^{\em} {\rm S} \right\}\,.
\label{E1:lowest}
\ee
Since the E1 operator is enhanced by $1/v$ relative to the leading M1 operator (\ref{M1:lowest}),
the recoil correction is of order $k_\gamma/m$ with respect to the leading term.
At relative order $v^2$, this correction is negligible for M1 allowed transitions ($k_\gamma \ll m v^2$),  
but should be considered for M1 hindered transitions, where $k_\gamma \sim m
v^2$.\footnote{
In the gauge non-invariant formulation of Appendix \ref{AppA}, the leading E1 operator 
is given by Eq.~(\ref{E1:lowestgd}). At relative order $v^2$, it contributes both to allowed and
hindered transitions.
}

The correction to the transition amplitude is given by
\bea
-~^{(1)}\bra{ n^{\prime\,1}S_0(-{\bf k}) \,\gamma({\bf k},\sigma)}
\int d^3R\, {\cal L}_{\rm E1}^{(0)} \; \ket{n^3S_1({\bf 0},\lambda)},  
\eea
where $\ket{ n^{\prime\,1}S_0(-{\bf k})}^{(1)}$ can be inferred from 
Eq.~(\ref{ketHPspin}). After a straightforward calculation, we obtain
\bea
\hspace{-5mm}
\overline{\cal A}\left[n^3S_1({\bf 0},\lambda) 
\to n^{\prime\,1}S_0(-{\bf k}) \gamma({\bf k},\sigma)\right]
 &=& 
\frac{k}{4m}\,\left( \delta_{n'n} + \frac{i}{3} \, \braQM{n'S} {\bf r}\cdot{\bf p}  \ketQM{nS} \right)
\,.
\label{ampl:S:9}
\eea

\subsection{Color-octet effects}
\label{wave:octet}
In Sec.~\ref{sec:fock}, we pointed out that a heavy quarkonium state also
contains higher Fock components, in particular, components made of a quark-antiquark pair in an octet configuration.
Color-octet effects are  regarded as one of the 
most distinctive benchmarks of NRQCD, and have been found to play a crucial role 
in several phenomenological applications, e.g. heavy quarkonium decays and productions \cite{Bodwin:1994jh}.
Indeed, color-octet effects are not included in any potential-model formulation and 
have not been considered so far in radiative transitions. A color-singlet 
quarkonium may develop a color-octet component 
by emitting and reabsorbing an ultrasoft gluon. A M1 transition may occur
either in the color-singlet or in the color-octet component.
If $\lQ \sim mv^2$, the process involving a singlet-octet-singlet transition 
is suppressed only by a factor $v^2$ with respect to the leading one,
and, therefore, relevant to our analysis.

In \cite{Brambilla:1999xf,Brambilla:1999xj}, the effect of octet components to the 
spectrum has been throughly investigated. The leading effect is given by\footnote{
Note that in \cite{Brambilla:1999xf,Brambilla:1999xj} a different normalization 
for the state $\ket{0}$ is used:  
$\ket{0} = \ket{0}_{\hbox{\cite{Brambilla:1999xf,Brambilla:1999xj}}}/\sqrt{N_c}$.}
\be
\delta E_{H(\lambda)} = \frac{i}{6}  \int_0^\infty \!\! dt \,
\bra{0} g{\bf E}^a({\bf R},0) \phi(0,t)^{\rm adj}_{ab} g{\bf E}^b({\bf R},t) \ket{0}\;
\braQM{H(\lambda)}
{\bf r} \, e^{-i(E_H^{(0)}-h_O^{(0)})t} \, {\bf r}
\ketQM{H(\lambda)}
\,,
\label{octet:energy:shift} 
\ee
where $\phi(0,t)^{\rm adj}_{ab}$ is a Wilson line in the 
adjoint representation connecting the point $({\bf R},0)$ to $({\bf R},t)$.
Note the appearance of the non-local condensate 
$\bra{0} g{\bf E}^a({\bf R},0) \phi(0,t)^{\rm adj}_{ab} g{\bf E}^b({\bf R},t) \ket{0}$, typical of the 
situation $\lQ\sim mv^2$. 
From Eq.~(\ref{ZH}) and (\ref{octet:energy:shift}) we may calculate the 
state normalization factor $\delta Z_H$:
\bea
\delta Z_{H(\lambda)} &=& \frac{\partial \delta E_{H(\lambda)}}{\partial E_H^{(0)}}
= \frac{1}{6}  \int_0^\infty \!\! dt\,t\, 
\bra{0} g{\bf E}^a({\bf R},0) \phi(0,t)^{\rm adj}_{ab} g{\bf E}^b({\bf R},t) \ket{0}\;
\nn\\
&& \qquad\qquad\qquad\qquad\qquad 
\times 
\braQM{H(\lambda)}
{\bf r} \, e^{-i(E_H^{(0)}-h_O^{(0)})t} \, {\bf r}\,
\ketQM{H(\lambda)}
\,.
\eea

\begin{figure}
\makebox[-15.5cm]{\phantom b}
\put(5,0){\epsfxsize=15truecm\epsffile{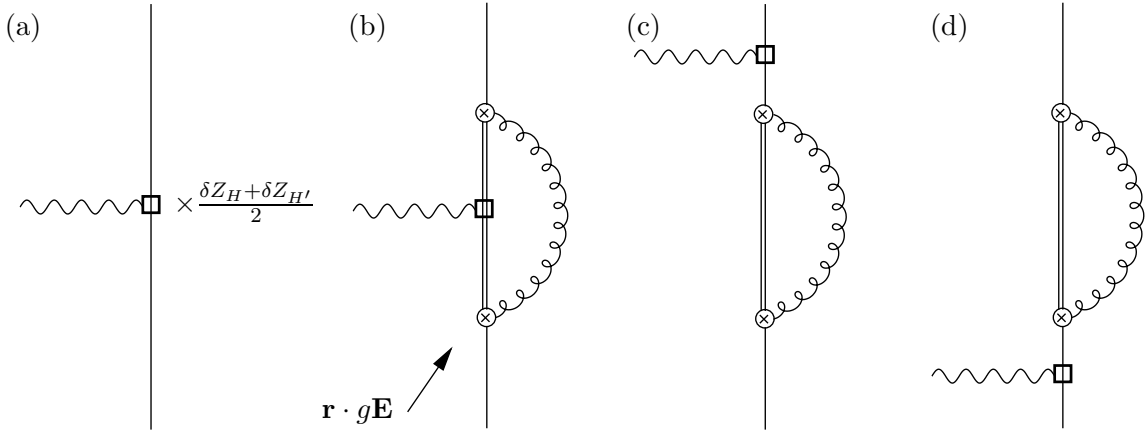}}
\put(0,150){(a)}
\put(130,150){(b)}
\put(235,150){(c)}
\put(350,150){(d)}
\put(120,5){${\bf r}\cdot g{\bf E}$}
\put(64,83){$\times  \frac{\delta Z_H + \delta Z_{H'}}{2}$}
\caption{
Octet contributions to M1 transitions.
The single and double lines represent the singlet and octet fields, respectively.
The circled cross stands for the vertex induced by the interaction (\ref{QCDE1}).}
\label{co:pNRQCD:M1:diagrams}
\end{figure}

The leading color-octet contribution is induced by the chromo-E1 operator (\ref{QCDE1}). 
At second order in ${\bf r}$, there are four diagrams contributing to the transition
amplitude, as shown in Fig.~\ref{co:pNRQCD:M1:diagrams}.
The contribution of Fig.~\ref{co:pNRQCD:M1:diagrams}(a) corresponds to 
\bea
&& \hspace{-8mm}
-^{(0)}\bra{ n^{\prime\,1}S_0({\bf P}^\prime) \,\gamma({\bf k},\sigma)}
\int d^3R\, {\cal L}_{\rm M1}^{(0)} \ket{n^3S_1({\bf 0},\lambda)}^{(2)}_\parallel 
\nn\\
&&\qquad\quad 
- ~_\parallel^{(2)}\bra{ n^{\prime\,1}S_0({\bf P}^\prime) \,\gamma({\bf k},\sigma)}
 \int d^3R\, {\cal L}_{\rm M1}^{(0)} \ket{n^3S_1({\bf 0},\lambda)}^{(0)}
\nn\\
& & \quad  
= (2\pi)^3\delta^3({\bf P}^\prime + {\bf k})\; 
\bar{\cal A}^{(0)}\left[n^3S_1({\bf 0},\lambda) \to n^{\prime\,1}S_0(-{\bf k}) \gamma({\bf k},\sigma)\right]
\nn\\
& & \qquad\quad 
\times 
c_F^\em \frac{\delta_{nn'} }{3}  \int_0^\infty \!\! dt\,t\, 
\bra{0} g{\bf E}^a({\bf R},0) \phi(0,t)^{\rm adj}_{ab} g{\bf E}^b({\bf R},t) \ket{0}\;
\nn\\
&& \qquad\quad 
\times \braQM{n'S} {\bf r} \, e^{-i(E_n^{(0)}-h_O^{(0)})t} \, {\bf r} \ketQM{nS}\,.
\label{octet:a}
\eea
This diagram only contributes to M1 allowed transitions.
The contribution of Fig.~\ref{co:pNRQCD:M1:diagrams}(b) corresponds to 
\bea
&& \hspace{-8mm}
-^{(1)}\bra{ n^{\prime\,1}S_0({\bf P}^\prime) \,\gamma({\bf k},\sigma)}
\int d^3R\, {\cal L}_{\rm M1}^{(0)\,{\rm octet}} \ket{n^3S_1({\bf 0},\lambda)}^{(1)}
\nn\\
& & \quad  
= (2\pi)^3\delta^3({\bf P}^\prime + {\bf k})\; 
\bar{\cal A}^{(0)}\left[n^3S_1({\bf 0},\lambda) \to n^{\prime\,1}S_0(-{\bf k}) \gamma({\bf k},\sigma)\right]
\nn\\
& & \qquad\quad 
\times 
c_F^\em \frac{i}{3}
\int_0^\infty \!\! dt \, 
\frac{e^{i(E^{(0)}_n-E^{(0)}_{n'})t}-1}{E^{(0)}_n-E^{(0)}_{n'}} 
\bra{0} g{\bf E}^a({\bf R},0) \phi(0,t)^{\rm adj}_{ab} g{\bf E}^b({\bf R},t) \ket{0}\;
\nn\\
&& \qquad\quad 
\times \braQM{n'S} {\bf r} \, e^{-i(E_n^{(0)}-h_O^{(0)})t} \, {\bf r} \ketQM{nS}\,,
\label{octet:b}
\eea
where $\displaystyle {\cal L}_{\rm M1}^{(0)\,{\rm octet}} = 
\int d^3r \;  {\rm Tr} \, \left\{
\frac{1}{2 m}
\; \left\{{\rm O}^\dagger , \bfsigma \cdot e e_Q {\bf B}^{\em}\right\} {\rm O} 
\right\}$. \\
The sum of the contributions of Figs.~\ref{co:pNRQCD:M1:diagrams}(c) and
\ref{co:pNRQCD:M1:diagrams}(d) gives:
\bea
&& \hspace{-8mm}
-^{(0)}\bra{ n^{\prime\,1}S_0({\bf P}^\prime) \,\gamma({\bf k},\sigma)}
 \int d^3R\, {\cal L}_{\rm M1}^{(0)} \ket{n^3S_1({\bf 0},\lambda)}^{(2)}_\perp
\nn\\
&&\qquad\quad 
- ^{(2)}_\perp\bra{ n^{\prime\,1}S_0({\bf P}^\prime) \,\gamma({\bf k},\sigma)}
 \int d^3R\, {\cal L}_{\rm M1}^{(0)} \ket{n^3S_1({\bf 0},\lambda)}^{(0)}
\nn\\
& & \quad  
= (2\pi)^3\delta^3({\bf P}^\prime + {\bf k})\; 
\bar{\cal A}^{(0)}\left[n^3S_1({\bf 0},\lambda) \to n^{\prime\,1}S_0(-{\bf k}) \gamma({\bf k},\sigma)\right]
\nn\\
& & \qquad\quad 
\times 
i c_F^\em \frac{1-\delta_{nn'} }{3}  \int_0^\infty \!\! dt\, 
\frac{1}{E^{(0)}_n-E^{(0)}_{n'}} 
\bra{0} g{\bf E}^a({\bf R},0) \phi(0,t)^{\rm adj}_{ab} g{\bf E}^b({\bf R},t) \ket{0}\;
\nn\\
&& \qquad\quad 
\times \braQM{n'S} {\bf r} \, 
\left(e^{-i(E_n^{(0)}-h_O^{(0)})t} - e^{-i(E_{n'}^{(0)}-h_O^{(0)})t}\right)
\, {\bf r} \ketQM{nS}\,.
\label{octet:cd}
\eea
These diagrams only contribute to M1 hindered transitions.

Remarkably the sum of all octet contributions at relative order $v^2$, 
i.e. Eqs.~(\ref{octet:a})-(\ref{octet:cd}), vanishes. This relies, again, 
on the fact that the leading M1 operator behaves like the identity operator in 
coordinate space. Our cancellation proof, indeed, 
very much resembles the $V^{\frac{\sigma\cdot B}{m}}_S$
non-renormalization proof given in Sec.~\ref{sec:VSB}. 
The cancellation of all octet contributions at relative order $v^2$ is
nothing more than a manifestation of the proper normalization of the
quarkonium state when higher-order Fock-space components are taken into
account.\footnote{On this see also Ref.~\cite{Voloshin:1979uv}.} 

Finally, we note that, since the leading operator responsible for 
E1 transitions, given in Eq.~(\ref{E1:lowest}), is not a unit operator in
coordinate space, the above cancellation mechanism does not apply there.
In the E1 case, when $\lQ \sim mv^2$, we will, in general, expect octet
corrections of the same size as the leading relativistic ones.
These have not been calculated so far, but may be of relevant phenomenological 
impact for electric-dipole transitions in weakly-coupled quarkonia.

\section{Final formulae incorporating all ${\cal O}(v^2)$ corrections}
\label{sec:final}
In this section, we sum all previously calculated contributions for
transition amplitudes and give expressions for the widths.

\subsection{${\cal O}(v^2)$ $S$-wave transition amplitude}
Summing Eqs.~(\ref{ampl:S:Lowestorder}), (\ref{ampl:S:1}),
(\ref{ampl:S:2}),  (\ref{ampl:S:3}),   (\ref{ampl:S:4}),
(\ref{ampl:S:5}),  (\ref{ampl:S:6}),  (\ref{ampl:S:7}),  (\ref{ampl:S:8}),
(\ref{ampl:S:9}),  we obtain the M1 transition amplitude with ${\cal O}(v^2)$
corrections included:
\bea
&&
\overline{\cal A}\left[n^3S_1({\bf 0},\lambda) 
\to n^{\prime\,1}S_0(-{\bf k}) \gamma({\bf k},\sigma)\right]
\nn\\
&& \hspace{15mm}
= \delta_{nn'} \,(1 + \kappa_Q^\em)
\nn\\
&& \hspace{18mm}
+ (1-\delta_{nn'})
\frac{1+\kappa^\em_Q}{m^2} \frac{\braQM{n'S} V^{\rm ss}({\bf r})\ketQM{nS}}{E^{(0)}_n-E^{(0)}_{n'}}
\nn\\
&& \hspace{18mm}
+ \braQM{n'S}
\left[
- \frac{1+\kappa^\em_Q}{24}\,k^2 {\bf r}^2 -  \left( \frac{5}{3}
+ \kappa^\em_Q\right) \frac{{\bf p}^2}{2m^2} +  \frac{\kappa^\em_Q}{6m}
rV_S^{(0)\,\prime} 
\right]
\ketQM{nS}
\,.
\label{ampl:S:total}
\eea
We have used $c_S^\em = 1+2\kappa_Q^\em$ and 
\be
i \frac{k}{m}\braQM{n'} {\bf r}\cdot {\bf p} \ketQM{n}
= \braQM{n'}
\left(
- 2 \frac{{\bf p}^2}{m^2} + \frac{rV_s^{(0)\,\prime}}{m}
\right)
\ketQM{n} + {\cal O}(v^4)\,,
\label{idQM}
\ee
which holds for 
$(1-\delta_{nn'})\, k \approx (1-\delta_{nn'})\, (E_n^{(0)} - E_{n'}^{(0)})
\sim (1-\delta_{nn'})\, mv^2$ and 
$\delta_{nn'}\, k \sim \delta_{nn'}\,mv^4$, which we neglect. 
Terms of the type $\delta_{nn'}\,k$, which are beyond 
our accuracy, have been neglected also in Eq.~(\ref{ampl:S:total}).

\subsection{$S$-wave transition widths}
Inserting (\ref{ampl:S:total}) into (\ref{transitionwidth}), we obtain the total
transition width for magnetic dipole transitions of $S$-wave quarkonium up to 
order $k_\gamma^3\,v^2/m^2$. We use the explicit one-loop value 
of $\kappa^\em_Q$ given in Eq.~(\ref{kQ}). In accordance to the power counting,  
we neglect order $\als^2(m)$ and order $\als(m)\, v^2$ corrections.
The final result reads 
\bea
\Gamma_{n^3S_1\to n^1S_0\,\gamma} &=& 
\frac{4}{3} \alpha e_Q^2 \frac{k_\gamma^3}{m^2}
\left[1 + C_F\frac{\als(m)}{\pi}
 -  \frac{5}{3}  \braQM{nS}\frac{{\bf p}^2}{m^2} \ketQM{nS} \right]\,,
\label{G31S}
\\
\Gamma_{n^3S_1\to n^{\prime\,1}S_0\,\gamma} 
\;&\lower4pt\vbox{\hbox{\rlap{\tiny \!\!\!\!$n\neq n'$}\lower-4pt\vbox{\hbox{$=$}}}}&\;
\frac{4}{3} \alpha e_Q^2 \frac{k_\gamma^3}{m^2}
\left[\braQM{n'S}
\left(
- \frac{k_\gamma^2 {\bf r}^2 }{24} -  \frac{5}{6} \frac{{\bf
    p}^2}{m^2} 
\right)
\ketQM{nS}
\right.
\nn\\
&&
\hspace{60mm}
\left.
+ \frac{1}{m^2} \frac{\braQM{n'S} V^{\rm ss}({\bf r})\ketQM{nS}}{E^{(0)}_n-E^{(0)}_{n'}}
\right]^2.
\nn\\
\label{G31Shin}
\eea
For completeness, we also give the $n^1S_0\to n^{\prime\,3}S_1\,\gamma$
transition width, which is relevant only for hindered transitions:
\bea
\Gamma_{n^1S_0\to n^{\prime\,3}S_1\,\gamma} 
\;&\lower4pt\vbox{\hbox{\rlap{\tiny \!\!\!\!$n\neq n'$}\lower-4pt\vbox{\hbox{$=$}}}}&\;
4 \alpha e_Q^2 \frac{k_\gamma^3}{m^2}
\left[\braQM{n'S}
\left(
- \frac{k_\gamma^2 {\bf r}^2}{24} -  \frac{5}{6} \frac{{\bf
    p}^2}{m^2} 
\right)
\ketQM{nS}
\right.
\nn\\
&&
\hspace{60mm}
\left.
- \frac{1}{m^2} \frac{\braQM{n'S} V^{\rm ss}({\bf r})\ketQM{nS}}{E^{(0)}_n-E^{(0)}_{n'}}
\right]^2.
\nn\\
\label{G13Shin}
\eea

Equations (\ref{G31S})-(\ref{G13Shin}) very much resemble those 
derived in \cite{Grotch:1984gf} and subsequently used in most 
non-relativistic potential model calculations of the magnetic dipole 
transitions in quarkonium (see, for instance, the review in
\cite{Brambilla:2004wf}). There are, however, some differences that we have 
already mentioned, but we would like to stress again.

{\bf (1)} Equations (\ref{G31S})-(\ref{G13Shin}) have a limited range of
validity that the EFT framework clarifies.
They are valid only in the weak-coupling limit, i.e. for quarkonia that fulfill 
the criterion $mv^2 \simg \lQ$.  The lowest bottomonium states and the
charmonium ground state may belong to quarkonia of this kind. As discussed in
Sec.~\ref{secstrong}, in the strong-coupling regime, i.e. for
higher-quarkonium excitations, at order $k_\gamma^3\,v^2/m^2$ more terms 
will, in principle, arise.

{\bf (2)} Equations (\ref{G31S})-(\ref{G13Shin}) do not contain contributions  
from a scalar interaction (proportional to $-\braQM{n'S} V^{\rm scalar}/m\ketQM{nS}$). 
This has been often used in potential models, but the analysis 
of Sec.~\ref{M1:assoc:LS:potential} has excluded such a contribution in pNRQCD.
 
{\bf (3)} The analysis in Sec.~\ref{sec:VSB} has also excluded (to all orders)
contributions to the quarkonium magnetic moment coming from the soft scale.
This allows us to substitute $\kappa^\em_Q$ with the value  
inherited from NRQCD, which at one loop is $\kappa^\em_Q = C_F\als/(2\pi)$.
The renormalization scale of $\als$ is $m$. 

{\bf (4)} In Sec.~\ref{wave:octet}, it has been shown that color-octet
contributions, not accessible to potential model analyses, cancel at order 
$k_\gamma^3\,v^2/m^2$. This leads to the conclusion that 
in the weak-coupling regime, at order $k_\gamma^3\,v^2/m^2$, M1 transitions 
are completely accessible to perturbation theory. In particular, once
the spin-spin potential is written at leading order in perturbation
theory (which is sufficient here),
\be
V^{\rm ss}({\bf r}) = \frac{8}{3} \, \pi \, C_F\als\, \delta^3({\bf r})\,,
\label{VssCoulomb}
\ee
and Eqs.~(\ref{G31S})-(\ref{G13Shin}) are calculated for 
Coulomb wave functions, the transition rates will only depend on the strong-coupling constant.

\subsection{$P$-wave transition widths}
\label{sec:final:P}
In this section, we consider only allowed M1 transitions between $P$-wave
states, since hindered $P$-wave transitions are unlikely to accommodate within
a weakly-coupled picture. The calculation proceeds very much like the 
analogous one for $S$-wave states, so we will not present details here. 
Octet contributions again cancel by the same argument as given for $S$-wave transitions.
At order $k_\gamma^3\,v^2/m^2$, only two operators contribute to M1 allowed transitions: 
$\displaystyle \frac{1}{4 m^3} \; \left\{ {\rm S}^\dagger , \bfsigma \cdot e e_Q {\bf
  B}^{\em} \right\} \bfnabla_r^2 {\rm S}$  and
$\displaystyle \frac{1}{4 m^2} \; \frac{r V_S^{(0)\,\prime}(r)}{2} 
\; \left\{{\rm S}^\dagger , \bfsigma\cdot\left[ \hat{\bf r} \times  \left(
  \hat{\bf r}\times e e_Q {\bf B}^{\em} \right) \right] \right\} {\rm S}$. 
Summing their contributions, at order $k_\gamma^3\,v^2/m^2$, the final results read
\bea
\Gamma_{n^3P_J\to n^1P_1\,\gamma} &=& 
\frac{4}{3} \alpha e_Q^2 \frac{k_\gamma^3}{m^2}
\left[1 + C_F\frac{\als(m)}{\pi}
- d_J\, \braQM{nP} \frac{{\bf p}^2}{m^2} \ketQM{nP} \right]\,,
\label{G31P}
\\
\Gamma_{n^1P_1\to n^3P_J\,\gamma} &=& (2J+1)
\frac{4}{9} \alpha e_Q^2 \frac{k_\gamma^3}{m^2}
\left[1 + C_F\frac{\als(m)}{\pi}
- d_J\, \braQM{nP} \frac{{\bf p}^2}{m^2} \ketQM{nP} \right]\,,
\label{G13P}
\eea
where $d_0= 1$, $d_1= 2$ and $d_2= 8/5$. 
We have made use of the virial theorem. 
Corrections induced by the operator $\displaystyle \frac{1}{4 m^2} \; \frac{r V_S^{(0)\,\prime}(r)}{2} 
\; \left\{{\rm S}^\dagger , \bfsigma\cdot\left[ \hat{\bf r} \times  \left(
  \hat{\bf r}\times e e_Q {\bf B}^{\em} \right) \right] \right\} {\rm S}$
vanish for $J=0$ states.

Combining Eq.~(\ref{G31S}) with Eqs.~(\ref{G31P}) and (\ref{G13P}), we obtain 
that, at leading order, the following relations hold:
\be
\frac{3\,\Gamma_{n^1P_1\to n^3P_0\,\gamma} - \Gamma_{n^3S_1\to n^1S_0\,\gamma}}{
\Gamma_{n^3P_2\to n^1P_1\,\gamma}- \Gamma_{n^3S_1\to n^1S_0\,\gamma}} = 10\,,
\qquad
\frac{\Gamma_{n^3S_1\to n^1S_0\,\gamma} - \Gamma_{n^1P_1\to n^3P_1\,\gamma}}{
\Gamma_{n^3P_2\to n^1P_1\,\gamma}- \Gamma_{n^3S_1\to n^1S_0\,\gamma}} = 5\,,
\label{ratiochib}
\ee
which follow from $\displaystyle \braQM{nS} \frac{{\bf p}^2}{m^2} \ketQM{nS} 
= \braQM{nP} \frac{{\bf p}^2}{m^2} \ketQM{nP} = - \frac{E_n^{(0)}}{m}$.\footnote{
So far, we have labeled $P$-wave states with their principal quantum number $n$.
In the next section, we will follow the usual convention for which a $\chi(1P)$
state is a $n=2$, $L=1$ state.  
}

\section{Applications}
\label{sec:phen}
We have remarked that Eqs.~(\ref{G31S})-(\ref{G13Shin}), (\ref{G31P}) and (\ref{G13P})
are valid only for weakly-coupled quarkonia. 
It is generally believed that the lowest-lying $b\overline{b}$ states,
$\Upsilon(1S)$ and $\eta_b$, are in the weak-coupling regime.
The situation for $\chi_b(1P)$, $h_b(1P)$, $\Upsilon(2S)$ and $\eta_b(2S)$
is more controversial, as it is for the lowest-lying $c\overline{c}$ states. 
We will assume that also these states are weakly coupled and see whether the
comparison between our predictions and the experimental data supports this
assumption or not. As for the $n=2$ charmonium states, it is undoubtedly inappropriate to consider them as 
weakly-coupled systems. A further complication of the $\psi(2S)$ and $\eta_c(2S)$ states is that they lie
too close to the open charm threshold, so that threshold effects should be
included in a proper EFT treatment. We will not consider them in our analysis. 
In the following, we shall apply  Eqs.~(\ref{G31S})-(\ref{G13Shin}) to 
$J/\psi \to \eta_c \,\gamma$, $\Upsilon(1S) \to \eta_b \,\gamma$,
$\Upsilon(2S) \to \eta_b(2S) \,\gamma$,  
$\Upsilon(2S) \to \eta_b \,\gamma$ and $\eta_b(2S) \to \Upsilon(1S) \,\gamma$, 
and Eqs.~(\ref{G31P}) and (\ref{G13P}) to 
$h_b(1P) \to \chi_{b0,1}(1P)\,\gamma$ and $\chi_{b2}(1P) \to h_b(1P) \,\gamma$.

\subsection{$J/\psi \to \eta_c \,\gamma$}
In potential models, the transition $J/\psi \to \eta_c \,\gamma$ 
has been often considered problematic to accommodate because 
its leading-order width is about $2.83$ keV (for $m_c = M_{J/\psi}/2 = 1548$
MeV), far away from the experimental 
value of  $(1.18 \pm 0.36)$ keV \cite{Eidelman:2004wy}.

Since we assume that the charmonium ground state is a weakly-coupled
quarkonium, Eq.~(\ref{G31S}) provides the transition width up to order $k_\gamma^3\,v^2_c/m^2$.
We may conveniently rewrite it as 
\bea
\Gamma_{J/\psi \to \eta_c \,\gamma} \!\! 
&=& \!\! 
\frac{16}{3} \alpha e_c^2 \frac{k_\gamma^3}{M_{J/\psi}^2}
\left[ 1 + C_F\frac{\als(M_{J/\psi}/2)}{\pi}
+\frac{2}{3}\frac{\braQM{1S} 
3 V_S^{(0)} - r V_S^{(0)\,\prime} 
\ketQM{1S}}{M_{J/\psi}}
\right]
\nn\\
\!\! 
&=& \!\! 
\frac{16}{3} \alpha e_c^2 \frac{k_\gamma^3}{M_{J/\psi}^2}
\left[ 1 + C_F\frac{\als(M_{J/\psi}/2)}{\pi} 
- \frac{2}{3} (C_F\als(p_{J/\psi}))^2
\right]
\,,
\label{Gamma1Sc}
\eea
where in the first line we have reexpressed the charm mass in terms of the $J/\psi$ mass,
$$
M_{J/\psi} = 2m_c + \braQM{1S} \frac{{\bf p}^2}{m_c} +  V_S^{(0)}(r)  \ketQM{1S},
$$
and made use of the virial theorem to get rid of the kinetic energy.
We have made explicit in Eq.~(\ref{Gamma1Sc}) that the normalization scale for the $\als$ 
inherited from $\kappa^\em_c$ is the charm mass ($\als(M_{J/\psi}/2) \approx 0.35$), 
and for the $\als$, which comes from the Coulomb potential, is the typical momentum 
transfer $p_{J/\psi} \approx m C_F \als(p_{J/\psi})/2  \approx 0.8$ GeV.
Numerically we obtain:
\be
\Gamma_{J/\psi \to \eta_c \,\gamma} =  (1.5 \pm 1.0)\;\hbox{keV}.
\label{Gamma1Scnum}
\ee
The uncertainty has been estimated by assuming the next corrections to be suppressed 
by a factor $\als^3(p_{J/\psi})$ with respect to the transition width in the non-relativistic limit.

Some comments are in order. First, we note that the uncertainty in (\ref{Gamma1Scnum}) is 
large. In our view, it fully accounts for the large uncertainty coming from
higher-order relativistic corrections, which may be large if we consider that 
those of order $k_\gamma^3\,v^2_c/m^2$ have reduced the leading-order result by about 50\%, and 
for the uncertainties in the normalization scales of the strong-coupling constant.
Both uncertainties may be only reduced by higher-order calculations. 

Despite the uncertainties, the value given in Eq.~(\ref{Gamma1Scnum}) is perfectly 
consistent with the experimental one. This means that assuming the ground-state charmonium 
to be a weakly-coupled system leads to relativistic corrections to the 
transition width of the right sign and size. This is not trivial.
If we look at the expression after the first equality in Eq.~(\ref{Gamma1Sc}), 
we may notice that $3 V_S^{(0)} - rV_S^{(0)\,\prime}$ is negative in the 
case of a Coulomb potential (i.e. it lowers the transition width), 
but positive in the case of a confining linear potential (i.e. it increases
the transition width). This may explain some of the difficulties met by potential 
models in reproducing $\Gamma_{J/\psi \to \eta_c \,\gamma}$. In any rate, it should be remembered that 
Eq.~(\ref{Gamma1Sc}) is not the correct expression to be used in the
strong-coupling regime.

\begin{figure}
\makebox[-16cm]{\phantom b}
\put(0,0){\epsfxsize=7.5truecm\epsffile{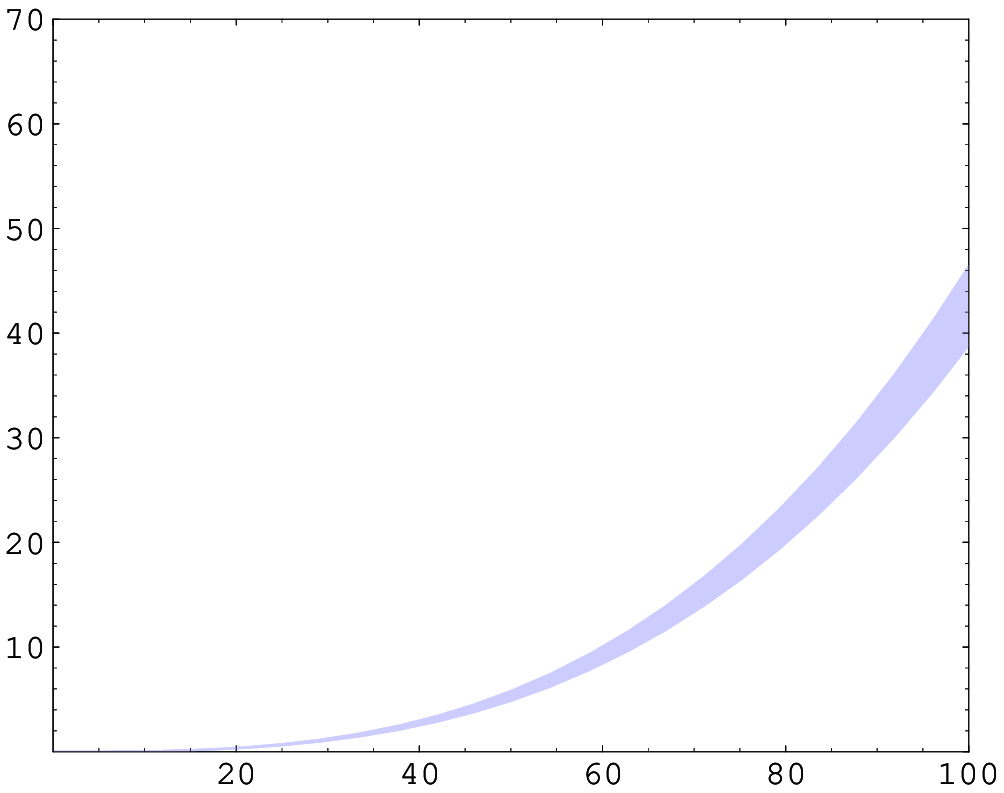}}
\put(250,0){\epsfxsize=7.5truecm\epsffile{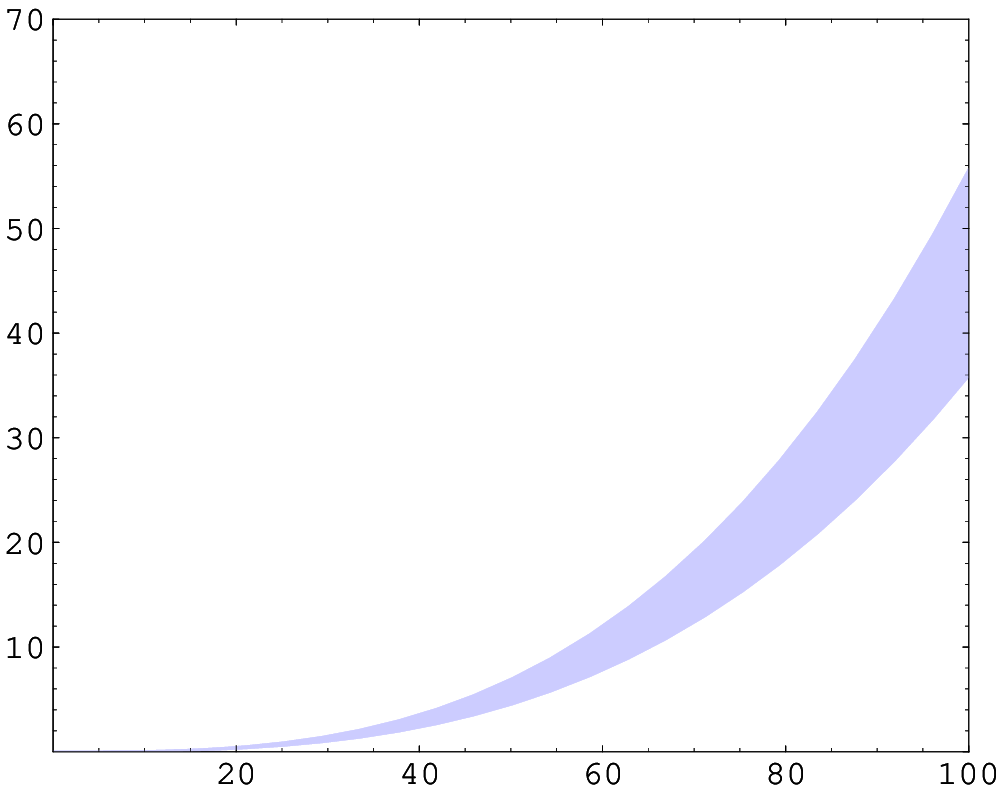}}
\put(160,-10){$k_\gamma$ (MeV)}
\put(410,-10){$k_\gamma$ (MeV)}
\put(20,150){$\Gamma_{\Upsilon(1S) \to \eta_b \,\gamma}$ (eV)}
\put(270,150){$\Gamma_{\Upsilon(2S) \to \eta_b(2S) \,\gamma}$ (eV)}
\caption{$\Gamma_{\Upsilon(1S) \to \eta_b \,\gamma}$ and $\Gamma_{\Upsilon(2S) \to \eta_b(2S) \,\gamma}$ 
as a function of the photon energy. 
}
\label{figMbb}
\end{figure}

\subsection{$\Upsilon(1S) \to \eta_b \,\gamma$, $\Upsilon(2S) \to \eta_b(2S) \,\gamma$}
\label{sec:app2S}
Allowed M1 transitions in the bottomonium system that may be treated 
by the weak-coupling formula (\ref{G31S}) are $\Upsilon(1S) \to \eta_b \,\gamma$ and, 
perhaps, $\Upsilon(2S) \to \eta_b(2S) \,\gamma$. We have
\bea
\Gamma_{\Upsilon(1S) \to \eta_b \,\gamma} &=& 
\frac{16}{3} \alpha e_b^2 \frac{k_\gamma^3}{M_{\Upsilon(1S)}^2}
\left[ 1 + C_F\frac{\als(M_{\Upsilon(1S)}/2)}{\pi} 
- \frac{2}{3} (C_F\als(p_{\Upsilon(1S)}))^2
\right]
\,,
\label{Gamma1Sb}
\\
\Gamma_{\Upsilon(2S) \to \eta_b(2S) \,\gamma} &=& 
\frac{16}{3} \alpha e_b^2 \frac{k_\gamma^3}{M_{\Upsilon(1S)}^2}
\left[ 1 + C_F\frac{\als(M_{\Upsilon(1S)}/2)}{\pi} 
- \left(\frac{C_F\als(p_{\Upsilon(1S)})}{2}\right)^2
\right.
\nn\\
&&
\hspace{50mm}
\left.
- \frac{5}{3}\left(\frac{C_F\als(p_{\Upsilon(2S)})}{4}\right)^2
\right]
\,,
\label{Gamma2Sb}
\eea
where we have expressed the $b$ mass in terms of the $\Upsilon(1S)$ mass.
We have made explicit that the renormalization scale for the $\als$, 
inherited from $\kappa^\em_b$, is the bottom mass ($\als(M_{\Upsilon(1S)}/2) \approx 0.22$), 
while for the $\als$, which comes from the Coulomb potential in the $\Upsilon(1S)$
system, is the typical momentum transfer $p_{\Upsilon(1S)} \approx m C_F \als(p_{\Upsilon(1S)})/2  \approx 1.2$ GeV, and 
for the $\als$, which comes from the Coulomb potential in the $\Upsilon(2S)$
system, is the typical momentum transfer $p_{\Upsilon(2S)} \approx m C_F \als(p_{\Upsilon(2S)})/4  \approx 0.9$ GeV. 

Since the $\eta_b$ has not been discovered yet, 
in Fig.~\ref{figMbb} we show $\Gamma_{\Upsilon(1S) \to \eta_b \,\gamma}$ 
and $\Gamma_{\Upsilon(2S) \to \eta_b(2S) \,\gamma}$ as a function of $k_\gamma$. 
The bands stand for the uncertainties calculated as 
the product of the transition widths in the non-relativistic limit by 
$\als^3(p_{\Upsilon(1S)})$ and $\als^3(p_{\Upsilon(2S)})$ respectively.
If we use the value of the $\eta_b$ mass given in \cite{Kniehl:2003ap}, i.e. $k_\gamma = 39\pm 13$ MeV, we have
\be 
\Gamma_{\Upsilon(1S) \to \eta_b \,\gamma} =  (3.6 \pm 2.9)\;\hbox{eV},
\ee 
which corresponds to a branching fraction of $(6.8 \pm 5.5)\times 10^{-5}$.

\begin{figure}
\makebox[-16cm]{\phantom b}
\put(0,0){\epsfxsize=7.5truecm\epsffile{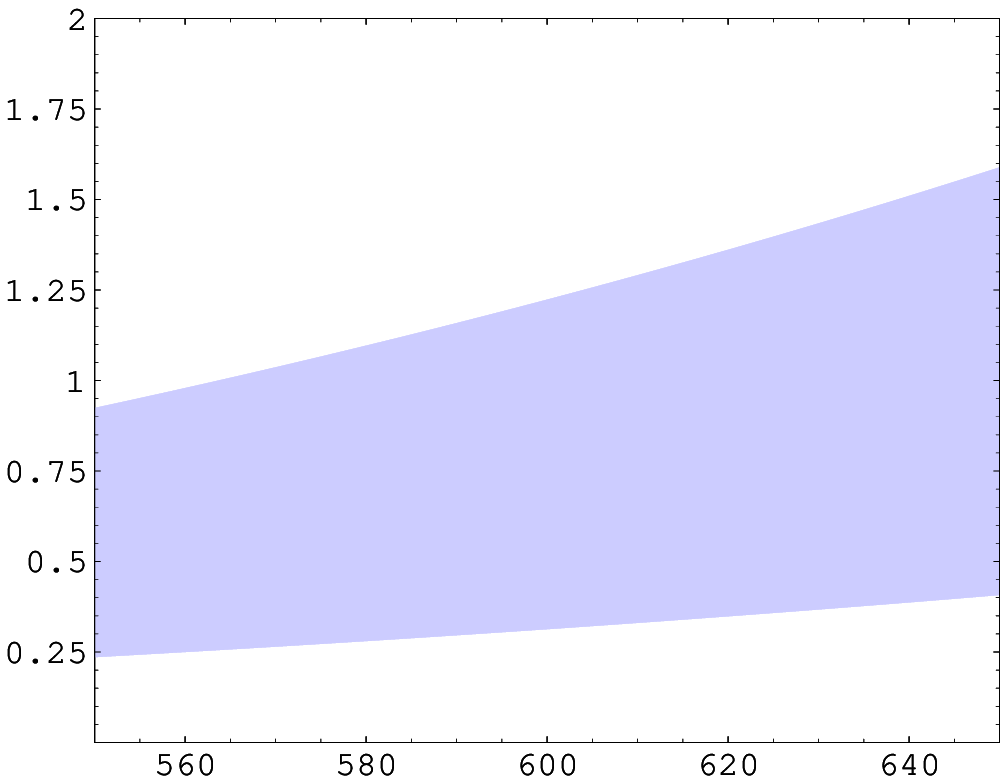}}
\put(250,0){\epsfxsize=7.5truecm\epsffile{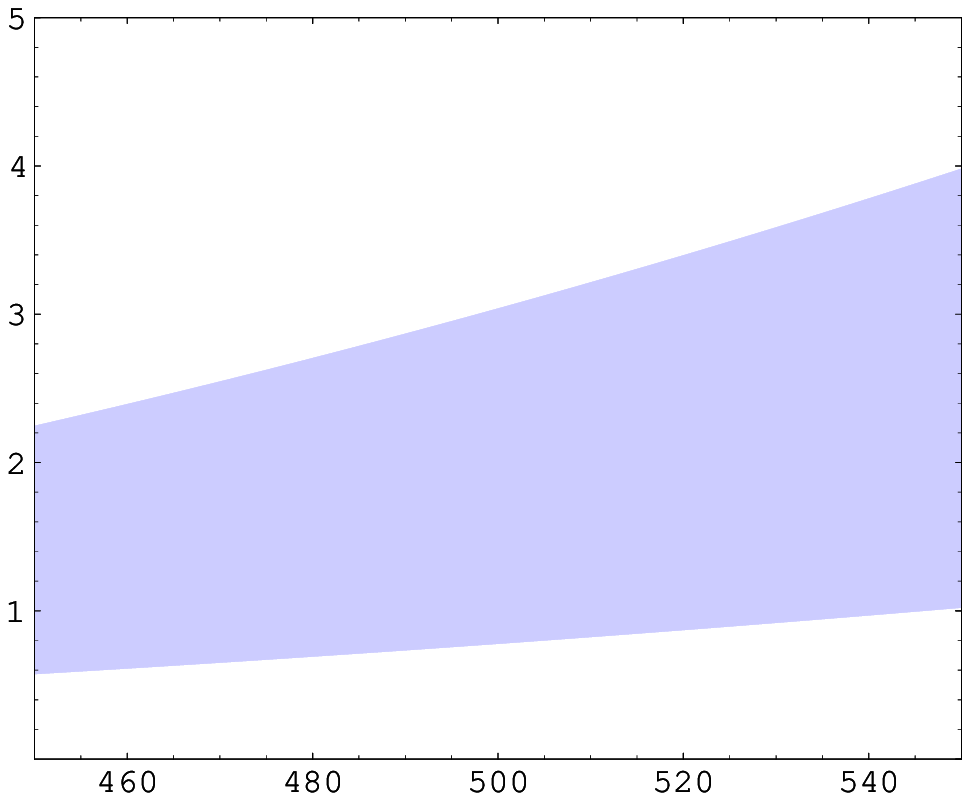}}
\put(160,-10){$k_\gamma$ (MeV)}
\put(410,-10){$k_\gamma$ (MeV)}
\put(25,150){$\Gamma_{\Upsilon(2S) \to \eta_b \,\gamma}$ (keV)}
\put(270,150){$\Gamma_{\eta_b(2S) \to \Upsilon(1S) \,\gamma}$ (keV)}
\caption{$\Gamma_{\Upsilon(2S) \to \eta_b \,\gamma}$ and $\Gamma_{\eta_b(2S) \to \Upsilon(1S) \,\gamma}$ 
as a function of the photon energy. 
}
\label{figMbbhin}
\end{figure}

\subsection{$\Upsilon(2S) \to \eta_b \,\gamma$, $\eta_b(2S) \to \Upsilon(1S) \,\gamma$}
For hindered M1 transitions, Eqs.~(\ref{G31Shin}) and (\ref{G13Shin}) only provide the leading-order expressions.
We consider, here, $\Upsilon(2S) \to \eta_b \,\gamma$ and $\eta_b(2S) \to \Upsilon(1S) \,\gamma$
transition widths that we write as
\bea
\Gamma_{\Upsilon(2S) \to \eta_b \,\gamma} &=& 
\frac{16}{3} \alpha e_b^2 \frac{k_\gamma^3}{M_{\Upsilon(1S)}^2}
\left[\frac{31\sqrt{2}}{81}(C_F\als)^2 
+ \frac{1024\sqrt{2}}{729}\frac{k_\gamma^2}{(M_{\Upsilon(1S)}C_F\als)^2}
\right]^2
\,,
\label{Gamma2Sbhin}
\\
\Gamma_{\eta_b(2S) \to \Upsilon(1S) \,\gamma} &=& 
16 \alpha e_b^2 \frac{k_\gamma^3}{M_{\Upsilon(1S)}^2}
\left[-\frac{41\sqrt{2}}{81}(C_F\als)^2 
+ \frac{1024\sqrt{2}}{729}\frac{k_\gamma^2}{(M_{\Upsilon(1S)}C_F\als)^2}
\right]^2
\,.
\label{Gamma2Sbhin13}
\eea
Since terms arising from the $1S$ and $2S$ system mix, 
it is difficult to assign a natural normalization scale to $\als$ appearing in 
Eqs.~(\ref{Gamma2Sbhin}) and (\ref{Gamma2Sbhin13}) without doing a higher-order calculation. 
In Fig.~\ref{figMbbhin}, we show a plot of $\Gamma_{\Upsilon(2S) \to \eta_b \,\gamma}$ 
and $\Gamma_{\eta_b(2S) \to \Upsilon(1S) \,\gamma}$ 
as a function of $k_\gamma$. The scale of $\als$ appearing in Eqs.~(\ref{Gamma2Sbhin}) and (\ref{Gamma2Sbhin13})
has been arbitrarily fixed to 1 GeV.
The bands stand for the uncertainties calculated as 
the products of the transition widths by $\als(p_{\Upsilon(2S)})$.

CLEO III recently set the 90\% upper limit for the branching fraction of
$\Upsilon(2S) \to \eta_b \,\gamma$ to be 0.5$\times 10^{-3}$ \cite{Artuso:2004fp}. 
The values plotted in Fig.~\ref{figMbbhin} are about a factor 10 
above the limit.\footnote{Large contributions stem from the spin-spin
  potential term. If instead of using 
$E_2^{(0)} - E_1^{(0)}$ in this term, we use the physical mass difference, the decay 
width reduces by about a factor one half.}  
Despite the fact that our calculation is just a leading order one 
and, therefore, potentially affected by large uncertainties, it 
is not obvious that perturbation theory may accommodate for such a large discrepancy.
In case, this may hint to a strongly-coupled interpretation of the bottomonium 
$2S$ states.\footnote{
A conclusion of this kind has been reached in \cite{Tormo:2005bs} from the
study of $\Upsilon(2S)\to X\, \gamma$ radiative decays.
On the other hand, the masses of the $n=2$ bottomonium states seem easier to
accommodate within a weakly-coupled picture \cite{Brambilla:2001fw,Brambilla:2004wu,Penin:2005eu}.}
\footnote{
It has been also argued that new physics may broaden the $\eta_b$ resonance, which thereby may have 
escaped detection at CLEO III \cite{Sanchis-Lozano:2005fj}.
}

\begin{figure}
\makebox[-16cm]{\phantom b}
\put(0,0){\epsfxsize=7.5truecm\epsffile{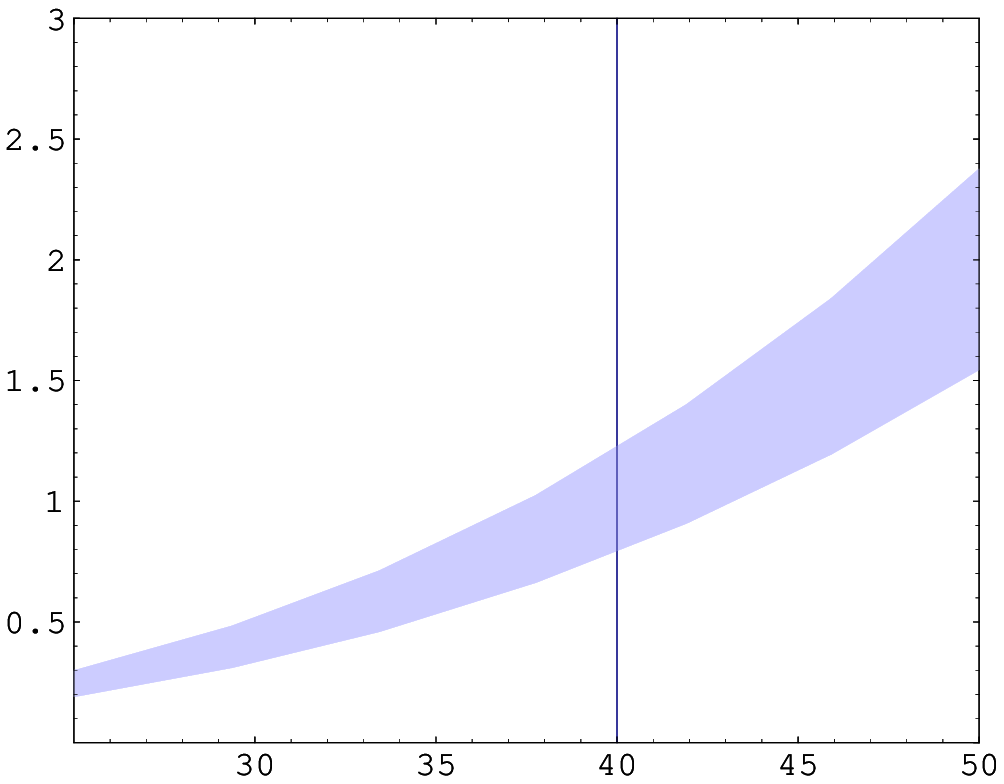}}
\put(250,0){\epsfxsize=7.5truecm\epsffile{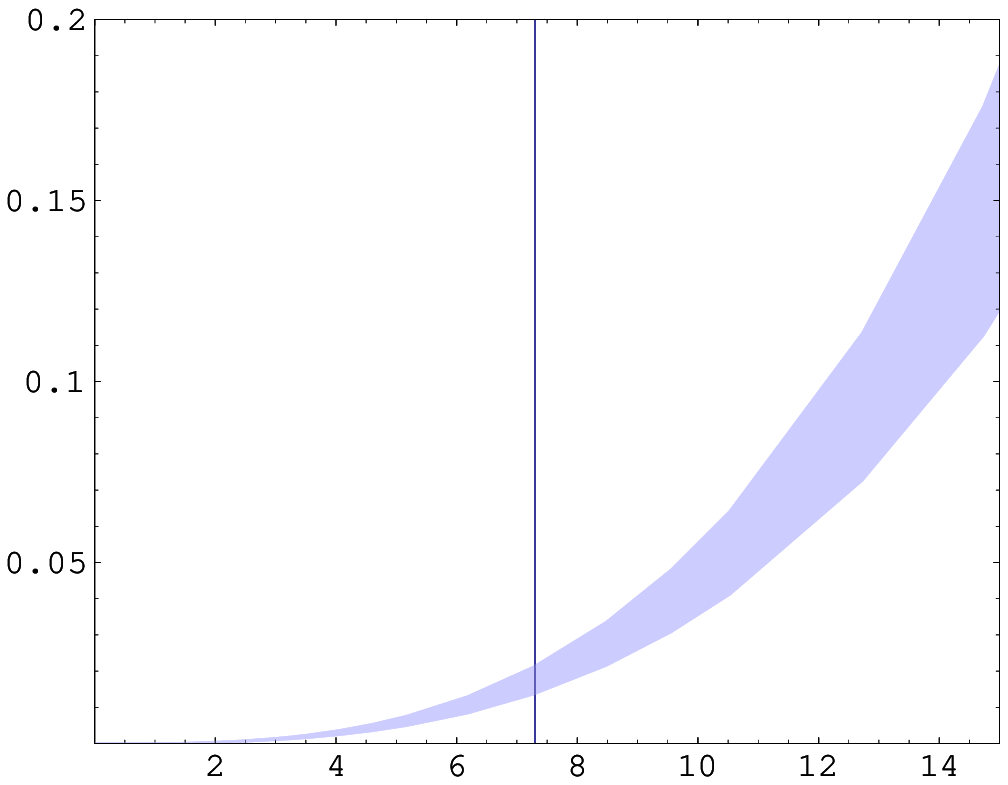}}
\put(125,-200){\epsfxsize=7.5truecm\epsffile{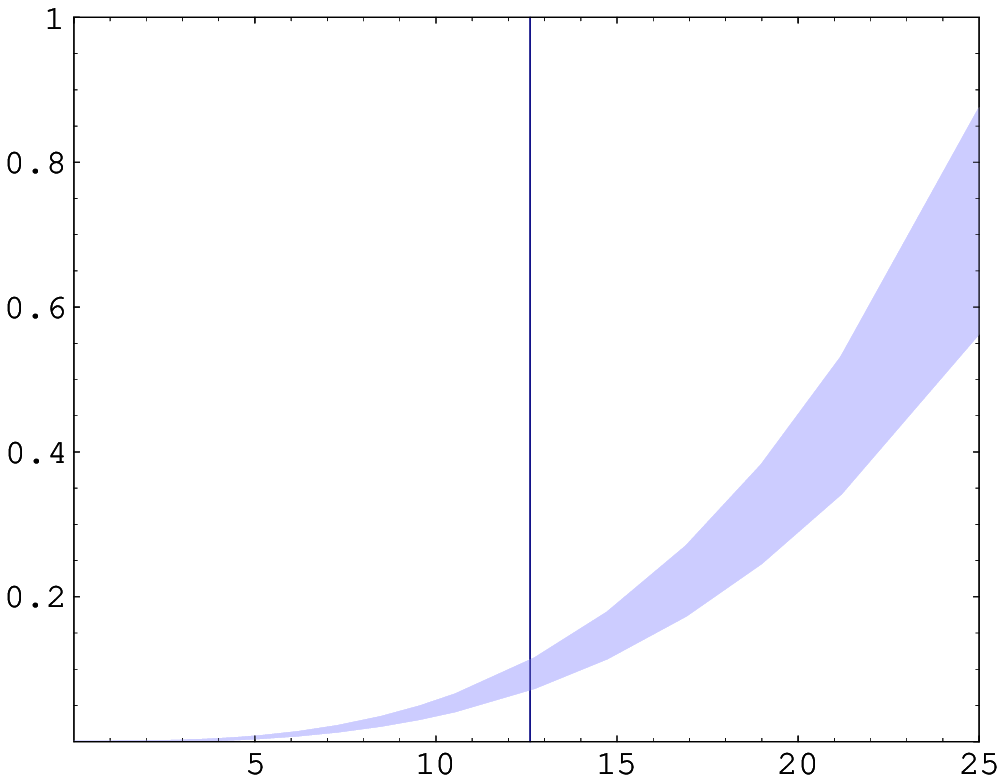}}
\put(160,-10){$k_\gamma$ (MeV)}
\put(410,-10){$k_\gamma$ (MeV)}
\put(285,-210){$k_\gamma$ (MeV)}
\put(25,150){$\Gamma_{h_b(1P) \to \chi_{b0}(1P) \,\gamma}$ ~~~~~~~~ (keV)}
\put(280,150){$\Gamma_{h_b(1P) \to \chi_{b1}(1P) \,\gamma}$ ~~~~~~~~ (keV)}
\put(150,-50){$\Gamma_{\chi_{b2}(1P) \to h_b(1P) \,\gamma}$ ~~~~~~~~ (keV)}
\caption{$\Gamma_{h_b(1P) \to \chi_{b0}(1P) \,\gamma}$,  
$\Gamma_{h_b(1P) \to \chi_{b1}(1P) \,\gamma}$ and  
$\Gamma_{\chi_{b2}(1P) \to h_b(1P) \,\gamma}$ as a function of the photon energy. 
The vertical lines correspond to the centre-of-gravity mass of the
$\chi_{bJ}(1P)$ states, which is about 9900 MeV. This is believed to be 
a rather accurate estimate of the $h_b(1P)$ mass. For this value of the $h_b(1P)$ mass, we obtain 
$\Gamma_{h_b(1P) \to \chi_{b0}(1P) \,\gamma} = 1 \pm 0.2$ keV, 
$\Gamma_{h_b(1P) \to \chi_{b1}(1P) \,\gamma} = 17 \pm 4$  eV and 
$\Gamma_{\chi_{b2}(1P) \to h_b(1P) \,\gamma} = 90 \pm 20$ eV.
}
\label{figMbb1P}
\end{figure}

\subsection{$h_b(1P) \to \chi_{b0,1}(1P)\,\gamma$, $\chi_{b2}(1P) \to h_b(1P) \,\gamma$}
$P$-wave M1 transitions that may be possibly described by 
pNRQCD in the weak-coupling regime are M1 allowed transitions between $n=2$ 
bottomonium states. Proceeding like in Sec.~\ref{sec:app2S}, we obtain
\bea 
\Gamma_{h_b(1P) \to \chi_{b0}(1P) \,\gamma} &=& 
\frac{16}{9} \alpha e_b^2 \frac{k_\gamma^3}{M_{\Upsilon(1S)}^2}
\left[ 1 + C_F\frac{\als(M_{\Upsilon(1S)}/2)}{\pi} 
- \left(\frac{C_F\als(p_{\Upsilon(1S)})}{2}\right)^2
\right.
\nn\\
&&
\hspace{50mm}
\left.
- \left(\frac{C_F\als(p_{\chi_b(1P)})}{4}\right)^2
\right]
\,,
\label{GammaP130}
\eea
\bea 
\Gamma_{h_b(1P) \to \chi_{b1}(1P) \,\gamma} &=& 
\frac{16}{3} \alpha e_b^2 \frac{k_\gamma^3}{M_{\Upsilon(1S)}^2}
\left[ 1 + C_F\frac{\als(M_{\Upsilon(1S)}/2)}{\pi} 
- \left(\frac{C_F\als(p_{\Upsilon(1S)})}{2}\right)^2
\right.
\nn\\
&&
\hspace{50mm}
\left.
- 2 \left(\frac{C_F\als(p_{\chi_b(1P)})}{4}\right)^2
\right]
\,,
\label{GammaP131}
\eea
and 
\bea 
\Gamma_{\chi_{b2}(1P) \to h_b(1P) \,\gamma} &=& 
\frac{16}{3} \alpha e_b^2 \frac{k_\gamma^3}{M_{\Upsilon(1S)}^2}
\left[ 1 + C_F\frac{\als(M_{\Upsilon(1S)}/2)}{\pi} 
- \left(\frac{C_F\als(p_{\Upsilon(1S)})}{2}\right)^2
\right.
\nn\\
&&
\hspace{50mm}
\left.
- \frac{8}{5}\left(\frac{C_F\als(p_{\chi_b(1P)})}{4}\right)^2
\right]
\,.
\label{GammaP321}
\eea

Since the $h_b(1P)$ has not been discovered yet, 
in Fig.~\ref{figMbb1P} we show 
$\Gamma_{h_b(1P) \to \chi_{b0}(1P) \,\gamma}$,  
$\Gamma_{h_b(1P) \to \chi_{b1}(1P) \,\gamma}$ and  
$\Gamma_{\chi_{b2}(1P) \to h_b(1P) \,\gamma}$ as a function of $k_\gamma$. 
We assume $p_{\Upsilon(\chi_b(1P))} \approx p_{\Upsilon(\Upsilon(2S))} \approx 0.9$ GeV. 
The bands stand for the uncertainties calculated as in Sec.~\ref{sec:app2S}.
Specific predictions of pNRQCD in the weak-coupling regime are 
also Eqs.~(\ref{ratiochib}) for $n=2$.

\section{Conclusion and Outlook}
\label{M1:summary}
The paper constitutes a thorough study of magnetic dipole transitions 
in the framework of non-relativistic EFTs of QCD and, 
in particular, of pNRQCD. The matching of the magnetic dipole operators 
at order $1/m$ and $1/m^2$ of pNRQCD has been performed to all orders in $\als$.
The matching at order $1/m^3$ has been carried out at leading order in the weak-coupling regime.
Relativistic corrections to the transition widths have been included in a systematic fashion.
Having achieved this, we could answer the questions raised in the introduction.
{\it (i)} The contribution to the quarkonium anomalous magnetic moment
coming from the soft scale vanishes to all orders. 
{\it (ii)} There are no contributions to the magnetic dipole operators
of the type induced by a scalar potential. {\it (iii)} In the weak-coupling 
regime, non-perturbative corrections due to color-octet contributions vanish at relative order $v^2$. 

Our final formulae (\ref{G31S})-(\ref{G13Shin}) are the same as in \cite{Grotch:1984gf}, once 
cleaned of the scalar potential and once the one-loop expression of the
quarkonium anomalous magnetic moment has been used. They are valid only for
quarkonia that fulfill the condition $mv^2 \simg \lQ$, i.e. only for the lowest-lying resonances.
The application of Eq.~(\ref{G31S}) to the transition $J/\psi\to
\eta_c\,\gamma$ shows that a weak-coupling treatment of the charmonium ground
state is consistent with the data. We also provide a prediction for the 
analogous transition in the bottomonium case.
Equations (\ref{G31P}) and (\ref{G13P}) are, to our knowledge, new.

Higher resonances that obey the condition $\lQ\sim mv$ are described by pNRQCD
in the strong-coupling regime. In this case, more operators, arising from the $1/m^3$
matching, are likely needed. 

This work provides a first step towards a complete treatment of quarkonium
radiative transitions in the framework of non-relativistic EFTs of QCD.
Some of the next steps are obvious and we shall conclude by commenting
on some of them.

{\bf (1)} To describe M1 transitions for higher resonances the completion of
the non-perturbative matching of the relevant pNRQCD operators at order
$1/m^3$ will be needed. The matching coefficients will be 
Wilson-loop amplitudes similar to those that describe the
non-perturbative potential at order $1/m^2$ \cite{Pineda:2000sz}.
This calculation, combined with a lattice simulation of the Wilson-loop amplitudes, may
provide a rigorous QCD derivation for all quarkonium M1 transitions below threshold. 

{\bf (2)} M1 hindered transitions of the type $\Upsilon(3S)\to \eta_b\gamma$
have been also studied at CLEO III \cite{Artuso:2004fp}. They involve emitted
photons whose momentum is comparable with the typical momentum-transfer in the 
bound state. Then one cannot rely anymore on the multipole expansion of the
external electromagnetic fields. In this case, however, one may exploit the hierarchy 
$p_{\eta_b} \gg \lQ \simg p_{\Upsilon(3S)}$. To our knowledge, this situation has not been analyzed so far.

{\bf (3)} Electric dipole  transitions have been mentioned only superficially in the paper.
The pNRQCD operators, relevant for E1 transitions beyond leading order, have not been
given. In the weak-coupling regime, octet contributions may be important and
can be worked out along the lines discussed here. However, most of E1
transitions may need to be treated in a strong-coupling framework.

\section*{Acknowledgments}
We thank Joan Soto for discussions and Tom Ferguson for correspondence.
A.V. acknowledges the financial support obtained inside the Italian 
MIUR program  ``incentivazione alla mobilit\`a di studiosi stranieri e 
italiani residenti all'estero''. A.V. and Y.J. were funded by 
Marie Curie Reintegration Grant contract MERG-CT-2004-510967.

\newpage

\appendix

\section{Final-state recoil effects}
\label{AppB}
We present here two alternative derivations of the final-state 
recoil effects calculated in Sec.~\ref{wave:recoil}.

\subsection{Recoil effects from Lorentz boosts}
The effect on the quarkonium state of higher-order potentials that depend on the   
centre-of-mass momentum ${\bf P}$ may also be calculated by boosting 
the quarkonium state at rest by $-{\bf P}/M_{H} \approx -{\bf P}/(2m)$ (i.e. minus the recoiling velocity).
The importance of boost effects on the final-state quarkonium  was first pointed out 
by Grotch and Sebastian in \cite{Grotch:1982bi}. In our language, their 
argument goes as follows.

The Lorentz-boost generators ${\bf K}$ of pNRQCD may be read from \cite{Brambilla:2003nt}.  
The leading spin-flipping contribution to ${\bf K}$ is given by 
\be
\delta {\bf K} = \int d^3 R \int d^3 r  \; 
\frac{i}{4m} {\rm Tr} \left\{ \left[ {\rm S}^\dagger, \bfsigma\times\bfnabla_r\right] 
{\rm S}\right\} \,.
\ee
It boosts the field ${\rm S}^\dagger$ by an amount 
\be
\delta {\rm S}^\dagger = -i \left[-\frac{{\bf P}}{2m}\cdot\delta{\bf K}, {\rm S}^\dagger\right]
= \epsilon_{ijk} \frac{{\bf P}^{i}}{8m^2}
\left\{ \bfnabla_r^k {\rm S}^\dagger, \bfsigma^j\right\} \,.
\label{deltaSdag}
\ee
Substituting (\ref{deltaSdag}) into
\be
\int d^3R \int d^3r \; e^{i{\bf P}\cdot {\bf R}} \, {\rm Tr} \left\{
\Phi_{H(\lambda)}({\bf r})\,\delta {\rm S}^\dagger({\bf r},{\bf R})\ket{0}\right\}\,,
\ee
we obtain Eq.~(\ref{ketHPspin}).

\subsection{Covariant formulation}
Final-state recoil effects are automatically included in any Lorentz covariant definition of the wave
function, like that one provided by the Bethe--Salpeter equation \cite{Salpeter:1951sz}.
In momentum space, the Bethe--Salpeter wave function has the following spin structure: 
\be
\Phi^{\rm BS}_H \propto \frac{{P\!\!\!\slash}/{2} + p \!\!\!\slash + m}{2m} \; 
\frac{1+ u\!\!\!\slash}{2} \; {\cal G}_H\; \frac{{P\!\!\!\slash}/{2} - p \!\!\!\slash -m}{2m} \,,
\ee
where, at the order we are interested in, $P^\mu \approx (2m,{\bf P})$ is the
centre-of-mass momentum, $p^\mu \approx (0,{\bf p})$ is the quark-antiquark relative
momentum, $u^\mu \approx (1,{\bf P}/(2m))$,  
${\cal G}_{n^1S_0}= \gamma^5$ and ${\cal G}_{n^3S_1(\lambda)}=
e_{n^3S_1}(\lambda) \hspace{-1.4cm} \slash \hspace{1.2cm}$, with $u \cdot e_{n^3S_1}(\lambda) = 0$.

Expanding $\Phi^{\rm BS}_H$ in ${\bf P}$ and ${\bf p}$ and keeping the upper-right $2\otimes 2$ block
we obtain
\bea 
\Phi^{\rm BS}_{n^1S_0}  &\propto&  
1 - \frac{i}{4m^2} {\bf P}\cdot (\bfsigma  \times {\bf p}) + \dots \,,
\\
\Phi^{\rm BS}_{n^3S_1(\lambda)}  &\propto&  
\bfsigma\cdot {\bf e} - \frac{i}{4m^2} {\bf P}\cdot ({\bf e}_{n^3S_1}(\lambda)
\times {\bf p}) + \dots \,.
\eea
The first terms in the equations give the spin structures of
Eqs.~(\ref{wavefun:jpsi}) and (\ref{wavefun:etac}), the second ones provide 
$\bra{0}S({\bf r},{\bf R})\ket{n^1S_0({\bf P})}^{(1)}$ and 
$\bra{0}S({\bf r},{\bf R})\ket{n^3S_1({\bf P},\lambda)}^{(1)}$ respectively, 
where $\ket{H({\bf P},\lambda)}^{(1)}$ has been given in  Eq.~(\ref{ketHPspin}).

\section{Gauge invariance}
\label{AppA}
In the main text, we have employed an explicitly gauge-invariant formulation.
In the literature, however, this has never been the case. 
As a consequence, partial results may differ.
In this appendix, in order to make contact with the existing literature, 
we recalculate M1 transitions in a formulation of pNRQCD where $U(1)_\em$ gauge 
invariance is not manifest at the Lagrangian level.
This means that we shall express the pNRQCD Lagrangian 
in terms of the fields ${\rm S}^\prime$ and ${\rm O}^\prime$
defined in Eq.~(\ref{Phi:to:S_and_O}). Of course, the final, 
total results are identical in the two formulations.

If the calculation of M1 transitions in pNRQCD is performed 
in terms of the field ${\rm S}^\prime$, there are two corresponding changes.

{\bf (1)} The first change concerns $1/m^2$ operators.
As discussed in Sec.~\ref{M1:assoc:LS:potential}, these 
may be obtained by projecting (\ref{NRQCDLS}) onto a two-quark state. 
If the projection is performed on (\ref{FockQQbar}) and (\ref{Phi:to:S_and_O}), 
we obtain the operator\footnote{
The leading operator in the multipole expansion, proportional to 
$\bfsigma\cdot \hat{\bf r} \times e e_Q {\bf A}^{\em}$, does not contribute 
to M1 transitions.}
\be
- \frac{1}{4 m^2}
\; \frac{r V_S^{(0)\,\prime}(r)}{2} 
\; \left\{{\rm S}^{\prime\,\dagger} , \bfsigma\cdot\left[ \hat{\bf r} \times  
  \left(\hat{\bf r}\cdot\bfnabla e e_Q {\bf A}^{\em}\right) \right] \right\} {\rm S}^\prime \,.
\label{Vm2LSgd}
\ee
It induces the following correction to $S$-wave transition widths
\bea
\hspace{-5mm}
\overline{\cal A}\left[n^3S_1({\bf 0},\lambda) 
\to n^{\prime\,1}S_0(-{\bf k}) \gamma({\bf k},\sigma)\right]
 &=& 
- \frac{1}{12m} \braQM{n'S} r  V_S^{(0)\,\prime} \ketQM{nS}
\,,
\label{ampl:S:7gd}
\eea
which differs by a factor $1/2$ from Eq.~(\ref{ampl:S:7}).

{\bf (2)} The second change concerns final-state recoil effects.
These have been calculated in a gauge invariant formulation 
in Sec.~\ref{wave:recoil}. In terms of the fields ${\rm S}^\prime$, E1 transitions 
are mediated by (to be compared with Eq.~(\ref{E1:lowest}))
\be
-2 i\int d^3 r \;  {\rm Tr} \, \left\{ 
{\rm S}^{\prime\,\dagger} \frac{ e e_Q  {\bf A}^{\em}\cdot \bfnabla_r}{m}
{\rm S}^\prime \right\}\,.
\label{E1:lowestgd}
\ee
The correction to $S$-wave transition widths induced by 
(\ref{E1:lowestgd}) on a recoiling final state is 
\bea
\hspace{-5mm}
\overline{\cal A}\left[n^3S_1({\bf 0},\lambda) 
\to n^{\prime\,1}S_0(-{\bf k}) \gamma({\bf k},\sigma)\right]
 &=& 
- \braQM{n'S} \frac{{\bf p}^2}{6m^2}  \ketQM{nS}
\,.
\label{ampl:S:9gd}
\eea
This is exactly the result first derived in \cite{Grotch:1982bi}.
Note that, at order $v^2$, Eq.~(\ref{ampl:S:9gd}) also contributes  
to M1 allowed transitions, while Eq.~(\ref{ampl:S:9}) only contributes  
to M1 hindered transitions.

Summing Eqs.~(\ref{ampl:S:7gd}) and (\ref{ampl:S:9gd}) we obtain
$$
\braQM{n'S} \left( - \frac{1}{12m} r  V_S^{(0)\,\prime} 
- \frac{{\bf p}^2}{6m^2} \right) \ketQM{nS}. 
$$
Summing Eqs.~(\ref{ampl:S:7}) and (\ref{ampl:S:9}) we obtain 
$$
\braQM{n'S} \left[ - \frac{1}{6m} r  V_S^{(0)\,\prime} 
+ \frac{k}{4m}\,\left( \delta_{n'n} + \frac{i}{3} \, {\bf r}\cdot{\bf p}  \right)\right]
\ketQM{nS}. 
$$
By using Eq.~(\ref{idQM}) and $\delta_{n'n} \, k \sim \delta_{n'n} \, m v^4$, one can easily see 
that the two expressions are equal at order $v^2$. It is straightforward to
perform the same check also in the case of $P$-wave transitions.


\begin{thebibliography}{99}
%\cite{Caswell:1985ui}
\bibitem{Caswell:1985ui}
W.~E.~Caswell and G.~P.~Lepage,
%``Effective Lagrangians For Bound State Problems In QED, QCD, And Other Field Theories,''
Phys.\ Lett.\ B {\bf 167}, 437 (1986).
%%CITATION = PHLTA,B167,437;%%

%\cite{Bodwin:1994jh}
\bibitem{Bodwin:1994jh}
G.~T.~Bodwin, E.~Braaten and G.~P.~Lepage,
%``Rigorous QCD analysis of inclusive annihilation and production of heavy
%quarkonium,''
Phys.\ Rev.\ D {\bf 51}, 1125 (1995)
[Erratum-ibid.\ D {\bf 55}, 5853 (1997)]
[hep-ph/9407339].
%%CITATION = HEP-PH 9407339;%%

%\cite{Pineda:1997bj}
\bibitem{Pineda:1997bj}
A.~Pineda and J.~Soto,
%``Effective field theory for ultrasoft momenta in NRQCD and NRQED,''
Nucl.\ Phys.\ Proc.\ Suppl.\  {\bf 64}, 428 (1998)
[arXiv:hep-ph/9707481].
%%CITATION = HEP-PH 9707481;%%

%\cite{Brambilla:1999xf}
\bibitem{Brambilla:1999xf}
N.~Brambilla, A.~Pineda, J.~Soto and A.~Vairo,
%``Potential NRQCD: An effective theory for heavy quarkonium,''
Nucl.\ Phys.\ B {\bf 566}, 275 (2000) [arXiv:hep-ph/9907240].
%%CITATION = HEP-PH 9907240;%%

%\cite{Brambilla:2004jw}
\bibitem{Brambilla:2004jw}
N.~Brambilla, A.~Pineda, J.~Soto and A.~Vairo,
%``Effective field theories for heavy quarkonium,''
Rev.\ Mod.\ Phys.\  {\bf 77}, 1423 (2005)
[arXiv:hep-ph/0410047].
%%CITATION = HEP-PH 0410047;%%

%\cite{Feinberg:1975hk}
\bibitem{Feinberg:1975hk}
G.~Feinberg and J.~Sucher,
%``Relativistic Calculation Of Radiative M1 Transitions In Charmonium,''
Phys.\ Rev.\ Lett.\  {\bf 35}, 1740 (1975).
%%CITATION = PRLTA,35,1740;%%

%\cite{Sucher:1978wq}
\bibitem{Sucher:1978wq}
J.~Sucher,
%``Magnetic Dipole Transitions In Atomic And Particle Physics,''
Rept.\ Prog.\ Phys.\  {\bf 41}, 1781 (1978).
%%CITATION = RPPHA,41,1781;%%

%\cite{Eichten:1978tg}
\bibitem{Eichten:1978tg}
E.~Eichten, K.~Gottfried, T.~Kinoshita, K.~D.~Lane and T.~M.~Yan,
%``Charmonium: The Model,''
Phys.\ Rev.\ D {\bf 17}, 3090 (1978)
[Erratum-ibid.\ D {\bf 21}, 313 (1980)];
%%CITATION = PHRVA,D17,3090;%%
%\cite{Eichten:1979ms}
%\bibitem{Eichten:1979ms}
E.~Eichten, K.~Gottfried, T.~Kinoshita, K.~D.~Lane and T.~M.~Yan,
%``Charmonium: Comparison With Experiment,''
Phys.\ Rev.\ D {\bf 21}, 203 (1980).
%%CITATION = PHRVA,D21,203;%%

%\cite{Kang:1978yw}
\bibitem{Kang:1978yw}
J.~S.~Kang and J.~Sucher,
%``Radiative M1 Transitions Of The Narrow Resonances,''
Phys.\ Rev.\ D {\bf 18}, 2698 (1978).
%%CITATION = PHRVA,D18,2698;%%

%\cite{Sebastian:1979gq}
\bibitem{Sebastian:1979gq}
K.~J.~Sebastian,
%``Relativistic Corrections To The Electric Dipole One Photon Transition Rates 
%Of Charmonium,''
Phys.\ Rev.\ D {\bf 26}, 2295 (1982).
%%CITATION = PHRVA,D26,2295;%%

%\cite{Karl:1980wm}
\bibitem{Karl:1980wm}
G.~Karl, S.~Meshkov and J.~L.~Rosner,
%``Quark Magnetic Moments And E1 Radiative Transitions In Charmonium,''
Phys.\ Rev.\ Lett.\  {\bf 45}, 215 (1980).
%%CITATION = PRLTA,45,215;%%

%\cite{Grotch:1982bi}
\bibitem{Grotch:1982bi}
H.~Grotch and K.~J.~Sebastian,
%``Magnetic Dipole Transitions Of Narrow Resonances,''
Phys.\ Rev.\ D {\bf 25}, 2944 (1982).
%%CITATION = PHRVA,D25,2944;%%

%\cite{Moxhay:1983vu}
\bibitem{Moxhay:1983vu}
P.~Moxhay and J.~L.~Rosner,
%``Relativistic Corrections In Quarkonium,''
Phys.\ Rev.\ D {\bf 28}, 1132 (1983).
%%CITATION = PHRVA,D28,1132;%%

%\cite{McClary:1983xw}
\bibitem{McClary:1983xw}
R.~McClary and N.~Byers,
%``Relativistic Effects In Heavy Quarkonium Spectroscopy,''
Phys.\ Rev.\ D {\bf 28}, 1692 (1983);
%%CITATION = PHRVA,D28,1692;%%
%\cite{Zambetakis:1983te}
%\bibitem{Zambetakis:1983te}
V.~Zambetakis and N.~Byers,
%``Magnetic Dipole Transitions In Onia,''
Phys.\ Rev.\ D {\bf 28}, 2908 (1983).
%%CITATION = PHRVA,D28,2908;%%

%\cite{Grotch:1984gf}
\bibitem{Grotch:1984gf}
H.~Grotch, D.~A.~Owen and K.~J.~Sebastian,
%``Relativistic Corrections To Radiative Transitions And Spectra Of Quarkonia,''
Phys.\ Rev.\ D {\bf 30}, 1924 (1984).
%%CITATION = PHRVA,D30,1924;%%

%\cite{Fayyazuddin:1993eb}
\bibitem{Fayyazuddin:1993eb}
Fayyazuddin and O.~H.~Mobarek,
%``E1 and M1 transitions in quarkonia,''
Phys.\ Rev.\ D {\bf 48}, 1220 (1993).
%%CITATION = PHRVA,D48,1220;%%

%\cite{Lahde:2002wj}
\bibitem{Lahde:2002wj}
T.~A.~Lahde,
%``Exchange current operators and electromagnetic dipole transitions in  heavy
%quarkonia,''
Nucl.\ Phys.\ A {\bf 714}, 183 (2003)
[arXiv:hep-ph/0208110].
%%CITATION = HEP-PH 0208110;%%

%\cite{Ebert:2002pp}
\bibitem{Ebert:2002pp}
D.~Ebert, R.~N.~Faustov and V.~O.~Galkin,
%``Properties of heavy quarkonia and B/c mesons in the relativistic quark
%model,''
Phys.\ Rev.\ D {\bf 67}, 014027 (2003)
[arXiv:hep-ph/0210381].
%%CITATION = HEP-PH 0210381;%%

%\cite{Barnes:2005pb}
\bibitem{Barnes:2005pb}
T.~Barnes, S.~Godfrey and E.~S.~Swanson,
%``Higher charmonia,''
Phys.\ Rev.\ D {\bf 72}, 054026 (2005)
[arXiv:hep-ph/0505002].
%%CITATION = HEP-PH 0505002;%%

%\cite{Khodjamirian:1979fa}
\bibitem{Khodjamirian:1979fa}
A.~Y.~Khodjamirian,
%``Dispersion Sum Rules For The Amplitudes Of Radiative Transitions In
%Quarkonium,''
Phys.\ Lett.\ B {\bf 90}, 460 (1980).
%%CITATION = PHLTA,B90,460;%%

%\cite{Brambilla:2004wf}
\bibitem{Brambilla:2004wf}
N.~Brambilla {\it et al.},
CERN-2005-005, (CERN, Geneva, 2005)
%``Heavy quarkonium physics,''
[arXiv:hep-ph/0412158].
%%CITATION = HEP-PH 0412158;%%

%\cite{LeYaouanc:1988fx}
\bibitem{LeYaouanc:1988fx}
A.~Le Yaouanc, L.~Oliver, O.~Pene and J.~C.~Raynal,
``Hadron Transitions In The Quark Model'', 
ed. Gordon and Breach (New York, 1988), 311p.
%\href{http://www.slac.stanford.edu/spires/find/hep/www?irn=1931024}{SPIRES entry}

%\cite{Manohar:1997qy}
\bibitem{Manohar:1997qy}
A.~V.~Manohar,
%``The HQET/NRQCD Lagrangian to order $\alpha/m^3$,''
Phys.\ Rev.\ D {\bf 56}, 230 (1997)
[hep-ph/9701294].
%%CITATION = HEP-PH 9701294;%%

%\cite{Beneke:1999zr}
\bibitem{Beneke:1999zr}
M.~Beneke,
%``Perturbative heavy quark-antiquark systems,''
arXiv:hep-ph/9911490.
%%CITATION = HEP-PH 9911490;%%

%\cite{Brambilla:2001q}
\bibitem{Brambilla:2001qk}
N.~Brambilla, Y.~Sumino and A.~Vairo,
%``Quarkonium spectroscopy and perturbative QCD: Massive quark loop
%effects,''
Phys.\ Rev.\ D {\bf 65}, 034001 (2002)
[arXiv:hep-ph/0108084].
%%CITATION = HEP-PH 0108084;%%

%\cite{Gromes:1984ma}
\bibitem{Gromes:1984ma}
D.~Gromes,
%``Spin Dependent Potentials In QCD And The Correct Long Range Spin Orbit Term,''
Z.\ Phys.\ C {\bf 26}, 401 (1984).
%%CITATION = ZEPYA,C26,401;%%

%\cite{Brambilla:2001xk}
\bibitem{Brambilla:2001xk}
N.~Brambilla, D.~Gromes and A.~Vairo,
%``Poincare invariance and the heavy-quark potential,''
Phys.\ Rev.\ D {\bf 64}, 076010 (2001)
[arXiv:hep-ph/0104068].
%%CITATION = HEP-PH 0104068;%%

%\cite{Brambilla:2003nt}
\bibitem{Brambilla:2003nt}
N.~Brambilla, D.~Gromes and A.~Vairo,
%``Poincare invariance constraints on NRQCD and potential NRQCD,''
Phys.\ Lett.\ B {\bf 576}, 314 (2003)
[hep-ph/0306107].
%%CITATION = HEP-PH 0306107;%%

%\cite{Eichten:1980mw}
\bibitem{Eichten:1980mw}
E.~Eichten and F.~Feinberg,
%``Spin Dependent Forces In QCD,''
Phys.\ Rev.\ D {\bf 23}, 2724 (1981).
%%CITATION = PHRVA,D23,2724;%%

%\cite{Pineda:2000sz}
\bibitem{Pineda:2000sz}
A.~Pineda and A.~Vairo,
%``The QCD potential at O(1/m**2): Complete spin-dependent and spin-independent result,''
Phys.\ Rev.\ D {\bf 63}, 054007 (2001)
[Erratum-ibid.\ D {\bf 64}, 039902 (2001)]
[arXiv:hep-ph/0009145].
%%CITATION = HEP-PH 0009145;%%

%\cite{Zhang:1991et}
\bibitem{Zhang:1991et}
X.~Zhang, K.~J.~Sebastian and H.~Grotch,
%``M1 decay rates of heavy quarkonia with a nonsingular potential,''
Phys.\ Rev.\ D {\bf 44}, 1606 (1991).
%%CITATION = PHRVA,D44,1606;%%

%\cite{Brambilla:2000gk}
\bibitem{Brambilla:2000gk}
N.~Brambilla, A.~Pineda, J.~Soto and A.~Vairo,
%``The QCD potential at O(1/m),''
Phys.\ Rev.\ D {\bf 63}, 014023 (2001)
[arXiv:hep-ph/0002250].
%%CITATION = HEP-PH 0002250;%%

%\cite{Brambilla:2003mu}
\bibitem{Brambilla:2003mu}
N.~Brambilla, A.~Pineda, J.~Soto and A.~Vairo,
%``The (m Lambda(QCD))**1/2 scale in heavy quarkonium,''
Phys.\ Lett.\ B {\bf 580}, 60 (2004)
[arXiv:hep-ph/0307159].
%%CITATION = HEP-PH 0307159;%%

%\cite{Brambilla:2001xy}
\bibitem{Brambilla:2001xy}
N.~Brambilla, D.~Eiras, A.~Pineda, J.~Soto and A.~Vairo,
%``New predictions for inclusive heavy-quarkonium P wave decays,''
Phys.\ Rev.\ Lett.\  {\bf 88}, 012003 (2002)
[arXiv:hep-ph/0109130].
%%CITATION = HEP-PH 0109130;%%

%\cite{Brambilla:2002nu}
\bibitem{Brambilla:2002nu}
N.~Brambilla, D.~Eiras, A.~Pineda, J.~Soto and A.~Vairo,
%``Inclusive decays of heavy quarkonium to light particles,''
Phys.\ Rev.\ D {\bf 67}, 034018 (2003)
[arXiv:hep-ph/0208019].
%%CITATION = HEP-PH 0208019;%%

%\cite{Brambilla:1999xj}
\bibitem{Brambilla:1999xj}
N.~Brambilla, A.~Pineda, J.~Soto and A.~Vairo,
%``The heavy quarkonium spectrum at order m alpha(s)**5 ln(alpha(s)),''
Phys.\ Lett.\ B {\bf 470}, 215 (1999)
[arXiv:hep-ph/9910238].
%%CITATION = HEP-PH 9910238;%%

%\cite{Voloshin:1979uv}
\bibitem{Voloshin:1979uv}
M.~B.~Voloshin,
%``Precoulombic Asymptotics For Energy Levels Of Heavy Quarkonium,''
Sov.\ J.\ Nucl.\ Phys.\  {\bf 36}, 143 (1982)
[Yad.\ Fiz.\  {\bf 36}, 247 (1982)].
%%CITATION = SJNCA,36,143;%%

%\cite{Eidelman:2004wy}
\bibitem{Eidelman:2004wy}
S.~Eidelman {\it et al.}  [Particle Data Group Collaboration],
%``Review of particle physics,''
Phys.\ Lett.\ B {\bf 592}, 1 (2004).
%%CITATION = PHLTA,B592,1;%%

%\cite{Kniehl:2003ap}
\bibitem{Kniehl:2003ap}
B.~A.~Kniehl, A.~A.~Penin, A.~Pineda, V.~A.~Smirnov and M.~Steinhauser,
%``M(eta/b) and alpha(s) from nonrelativistic renormalization group,''
Phys.\ Rev.\ Lett.\  {\bf 92}, 242001 (2004)
[arXiv:hep-ph/0312086].
%%CITATION = HEP-PH 0312086;%%

%\cite{Artuso:2004fp}
\bibitem{Artuso:2004fp}
M.~Artuso {\it et al.}  [CLEO Collaboration],
%``Photon transitions in Upsilon(2S) and Upsilon(3S) decays,''
Phys.\ Rev.\ Lett.\  {\bf 94}, 032001 (2005)
[hep-ex/0411068].
%%CITATION = HEP-EX 0411068;%%

%\cite{Tormo:2005bs}
\bibitem{Tormo:2005bs}
X.~G.~i.~Tormo and J.~Soto,
%``Radiative decays and the nature of heavy quarkonia,''
arXiv:hep-ph/0511167.
%%CITATION = HEP-PH 0511167;%%

%\cite{Brambilla:2001fw}
\bibitem{Brambilla:2001fw}
N.~Brambilla, Y.~Sumino and A.~Vairo,
%``Quarkonium spectroscopy and perturbative QCD: A new perspective,''
Phys.\ Lett.\ B {\bf 513}, 381 (2001)
[arXiv:hep-ph/0101305].
%%CITATION = HEP-PH 0101305;%%

%\cite{Brambilla:2004wu}
\bibitem{Brambilla:2004wu}
N.~Brambilla and A.~Vairo,
%``The 1P quarkonium fine splittings at NLO,''
Phys.\ Rev.\ D {\bf 71}, 034020 (2005)
[arXiv:hep-ph/0411156].
%%CITATION = HEP-PH 0411156;%%

%\cite{Penin:2005eu}
\bibitem{Penin:2005eu}
A.~A.~Penin, V.~A.~Smirnov and M.~Steinhauser,
%``Heavy quarkonium spectrum and production / annihilation rates to order
%beta(0)**3 alpha(s)**3,''
Nucl.\ Phys.\ B {\bf 716}, 303 (2005)
[arXiv:hep-ph/0501042].
%%CITATION = HEP-PH 0501042;%%

%\cite{Sanchis-Lozano:2005fj}
\bibitem{Sanchis-Lozano:2005fj}
M.~A.~Sanchis-Lozano,
%``Prospects of probing new physics in bottomonium decays and spectroscopy,''
arXiv:hep-ph/0503266.
%%CITATION = HEP-PH 0503266;%%

%\cite{Salpeter:1951sz}
\bibitem{Salpeter:1951sz}
E.~E.~Salpeter and H.~A.~Bethe,
%``A Relativistic Equation For Bound State Problems,''
Phys.\ Rev.\  {\bf 84}, 1232 (1951).
%%CITATION = PHRVA,84,1232;%%
\end{thebibliography}
\end{document}